\documentclass[11pt]{article}

\usepackage[letterpaper,margin=1in]{geometry}
\usepackage[utf8]{inputenc}
\usepackage[T1]{fontenc}
\usepackage{lmodern}
\usepackage{microtype}
\usepackage{amsmath,amssymb}
\usepackage{mathtools}
\usepackage{graphicx}
\usepackage{xcolor}
\usepackage{subcaption}
\usepackage{algorithm}
\usepackage{algpseudocode}
\usepackage{url}
\usepackage[numbers,sort&compress]{natbib}
\usepackage{authblk}
\usepackage[hidelinks]{hyperref}

\begin{document}

\title{The coupled spread of information  and potential-driven motion in evacuating agent populations}

\author[1,\textdagger]{Benjamin A. Jasperson}
\author[1,\textdagger]{Logan Dehay}
\author[1]{Rahul Gulati}
\author[2]{Siddhartha Srivastava}
\author[1,*]{Krishna Garikipati}

\affil[1]{Department of Aerospace and Mechanical Engineering, University of Southern California, Los Angeles, California, USA}
\affil[2]{Department of Aerospace Engineering, Auburn University, Auburn, Alabama, USA}

\date{\small
  \textsuperscript{\textdagger}These authors contributed equally to this work.\\
  \textsuperscript{*}Corresponding author: \href{mailto:garikipa@usc.edu}{garikipa@usc.edu}%
}

\maketitle

\begin{abstract}
Whether a population evacuates safely depends on two coupled processes: how a warning spreads through it, and how that information moves people.
We model this coupling with an agent-based framework, in which each agent's information state---susceptible, exposed, informed or reverted---gates its motion.
Only informed agents respond to a global background potential encoding hazards and safe zones, while all agents interact locally through a Lennard-Jones potential. The total potential experienced by an agent drives an It\^{o} stochastic differential equation.
Tests show that landscapes with potential minima leave far fewer agents never exposed than in a flat, uniform potential control landscape, as informed agents driven toward the minima sweep past susceptible neighbors.
In our tests, conduit routes raise the fraction of agents in the safe-zone minima, but leave greater numbers of agents never exposed, when compared to an unguided evacuation.
This is attributed to conduits shortening trips at the cost of informed agents having fewer encounters with susceptible neighbors who could receive their information.
Independently, the effective information transmission rate estimated from these encounters falls roughly sixty-fold below its well-mixed value within the first time unit, because state transitions create spatial mechanisms of net transmission suppression  that mean-field closure does not capture.
\end{abstract}


\section{Introduction}

When a warning is issued, people seldom act on it immediately.
They must first receive it, interpret it, and often seek confirmation from others before deciding to move \cite{lindellProtectiveActionDecision2012, kuligowskiProcessHumanBehavior2009}.
Many learn of a threat from neighbors or family rather than from official channels \cite{sorensenHazardWarningSystems2000}, and the resulting delays before movement are long and strongly right-skewed \cite{purserQuantificationBehaviourEngineering2001, lovreglioModelDecisionmakingProcess2015, woodMillingPublicWarnings2018}. %
Who reaches safety, and when, therefore depends on two coupled processes: the spread of information through a population, and the movement that information triggers.

Existing models tend to capture one of these processes at a time.
Compartmental models, descending from the work of Kermack and McKendrick \cite{kermack1927contribution,kermack1932contributions}, track individuals through states such as susceptible, exposed, informed and reverted.
In their classical form they assume a well-mixed population, and have been applied to rumor and information spread \cite{MakiThompson1973}.
Spatial extensions let states change through local pair interactions while individuals diffuse \cite[Sec.~10.5]{helbingQuantitativeSociodynamicsStochastic2010}.
Others recast the dynamics as hydrodynamic descriptions of social mobility, in which each group moves up the gradient of its own utility \cite{searaSociohydrodynamicsDatadrivenModeling2025}.
Both describe coarse-grained densities rather than individual agents, and neither lets a change of state trigger directed motion.
In Helbing's diffusion model, location and opinion changes occur independently, and even where mobility depends on opinion, state changes do not direct motion toward a goal \cite{helbingQuantitativeSociodynamicsStochastic2010}.
In that of Seara et al., group membership is fixed \cite{searaSociohydrodynamicsDatadrivenModeling2025}.
Force-based crowd models, in contrast, treat pedestrians as particles driven by social forces \cite{helbingSocialForceModel1995, helbingSimulatingDynamicalFeatures2000, makmulSocialForceModel2020}, reproducing collective phenomena from simple individual rules \cite{moussaidHowSimpleRules2011}.
Evacuation models of this kind commonly assign each agent a pre-movement time drawn independently of the agent's surroundings, rather than arising from contact with informed neighbors, and models of contagion among mobile agents \cite{peruaniDynamicsSteadyStates2008} %
couple motion and spread but let motion proceed independently of the information itself.
Full coupling can be bi-directional: an informed agent moves toward safety, and in doing so becomes unable to warn those it leaves behind---the reverse of the situation in many animal groups, where a small informed subset guides an uninformed majority \cite{couzinEffectiveLeadershipDecisionmaking2005}. %
Whether the departure of informed individuals leaves others behind, and how the environment shapes that outcome, cannot be addressed by a model that treats information and motion separately.

Here we develop an agent-based framework that bi-directionally couples the spread of information and motion.
Each agent is in a discrete information state: susceptible to information ($S$), exposed to information ($E$), informed ($I$) and reverted ($R$) from $I$.  Agents progress through these states as $S \to E \to I \to R$, with Gamma-distributed delays that represent heterogeneous pre-movement behavior.
Agent motion is governed by It\^{o} stochastic differential equations whose drift is the negative gradient of a the total potential, extending the stochastic motion used in agent-based models of disease spread among mobile individuals \cite{alexiadisSimulationPandemicsReal2021, horiiCalibrationVerificationStochastic2024}.
The potential plays a role similar to the static floor field of cellular-automaton pedestrian models \cite{bursteddeSimulationPedestrianDynamics2001} and to the artificial potential fields used in robotic path planning \cite{volpeManipulatorControlSuperquadric1990, hwangPotentialFieldApproach1992}.
Hazards are represented as potential maxima, safe zones as minima, and evacuation routes as curvilinear, sloping valleys.
Only informed agents respond to the environmental potential, so an agent's information state determines its motion, while agents' positions determine, through finite interaction radii, who can be exposed to an informed agent.
Because the potential is continuous and grid-free, the dynamics remain in a form suited to later coarse-graining.

Using this framework, we simulate evacuations over landscapes with and without conduits, for populations with homogeneous and multimodal response delays.
Landscapes in which informed agents respond to the environmental potential leave far fewer agents never exposed than an otherwise identical flat control landscape, because moving informed agents sweep past susceptible neighbors they would not otherwise encounter.
Evacuation conduits raise the fraction of agents that reach the safe zone but leave more agents never exposed than unguided evacuation.
This is attributed to a faster, narrower route reducing how many susceptible agents an informed agent sweeps past before it reverts.
Two further results hold across the landscapes studied.
The number of agents left in the exposed state is set by the fraction of the population with very long delays, whereas the number never exposed depends on how agents move.
And at the contact densities studied, the effective transmission rate falls roughly sixty-fold below its well-mixed value within the first time unit, because transmission itself creates spatial mechanisms of net transmission suppression that a mean-field closure with a constant, or state-dependent, transmission rate cannot represent.

The remainder of this paper is as follows.
Section~\ref{sec:comp-framework} introduces the model, Section~\ref{sec:appl_numer} presents the evacuation studies, and Section~\ref{sec:disc} discusses the limits of mean-field descriptions and the implications for evacuation planning.
Sensitivity studies and simulation details appear in the appendices.

\section{Computational framework and microscopic mechanics}
\label{sec:comp-framework}

\begin{figure}[tbp]
    \centering
    \begin{subfigure}[t]{0.3\linewidth}
        \includegraphics[width=\linewidth]{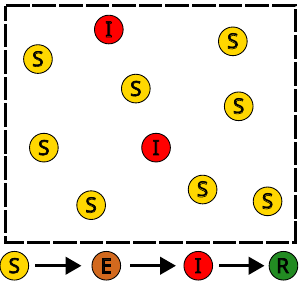}
        \caption{}
        \label{fig:domain_viz}
    \end{subfigure}
    \begin{subfigure}[t]{0.3\linewidth}
        \includegraphics[width =\linewidth]{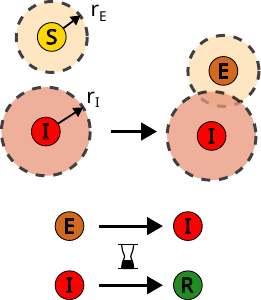}
        \caption{}
        \label{fig:re_ri_overlap}
    \end{subfigure}
    \begin{subfigure}[t]{0.3\linewidth}
        \includegraphics[width=\linewidth]{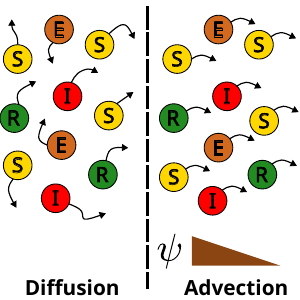}
        \caption{}
        \label{fig:adv_diff_viz}
    \end{subfigure}
    \caption{
    (a) Schematic of simulation domain, initialized with $N$ agents in one of four states ($S$, $E$, $I$, $R$).
    (b) Illustration of the spread of information through an agent population by overlapping spheres of interaction of individual agents.
    Each agent's radius of interaction varies according to their current state.
    (c) Illustration of the drift and stochastic components of individual agents' motions.
    }
    \label{fig:advectiondiffusion}
\end{figure}

Consider a domain $\Omega = (0,l)^2$ with $N$ agents that can each occupy one of the following states: susceptible, exposed, informed and reverted (Fig. \ref{fig:domain_viz}).
The number of agents in each state is denoted by $S, E, I, R$ respectively, satisfying $S + E + I + R = N$.
The corresponding fractions are: $s = S/N$, $e = E/N$, $i = I/N$ and $r = R/N$.

\subsection{The state dynamics of information exchange}

Each of the agents has an internal clock with elapsed time $t_c$ that begins at birth.
If all agents are initialized at the start of the process, then $t_c$ will equal the process time, $t$.
Additionally, each agent is initialized with four internal properties: an exposure cutoff radius, $r_E$, reflecting its susceptibility to be exposed to new information; an exposure time lag, $\tau_E$, representing the temporal delay before transitioning to the informed state; an information transmission cutoff radius, $r_I$, representing its sphere of influence; and a reversion time lag, $\tau_I$, dictating how long it retains the information.

The agent state update logic is detailed in Algorithm \ref{alg:state_update}.
Information transfer initiates when a susceptible ($S$) agent's ball of exposure with radius $r_E$ overlaps with an informed ($I$) agent's ball of information transmission with radius $r_I$ (i.e., their spatial separation is less than $r_E + r_I$, Fig. \ref{fig:re_ri_overlap}).
Upon overlap, the susceptible agent immediately becomes exposed ($E$).
After the time lag $\tau_E$, the exposed agent transitions to the informed state ($I$).
This informed agent then actively transfers information to other susceptible agents until it transitions to the reverted state ($R$) after its characteristic time lag $\tau_I$.
This process is tracked by each agent's transition timestamps relative to the simulation start $t_0$: initial exposure ($t_e$), transition to informed ($t_i$), and transition to reverted ($t_r$).
The individual residence durations can be computed as $\Delta t_e = t_i - t_e$ and $\Delta t_i = t_r - t_i$.

\begin{algorithm}[tbp]
\caption{Adjust State of an Agent}
\begin{algorithmic}[1]
\Procedure{AdjustState}{}
    \State $t_c \gets t_c + \Delta t$
    \If{agent is in exposed neighborhood \& agent is susceptible}
        \State $t_e \gets t_c$
        \State agent $\gets$ exposed \Comment{S$\rightarrow$E}
    \EndIf
    \If{agent is exposed}
        \If{$t_c-t_e >  \tau_E$}
            \State agent $\gets$ informed
            \State $t_i \gets t_c$ \Comment{E$\rightarrow$I}
        \EndIf
    \EndIf
    \If{agent is informed}
        \If{$t_c - t_i > \tau_I$}
            \State agent $\gets$ reverted
            \State $t_r \gets t_c$ \Comment{I$\rightarrow$R}
        \EndIf
    \EndIf
\EndProcedure
\end{algorithmic}
\label{alg:state_update}
\end{algorithm}

Heterogeneity is introduced to the population scale temporal behavior if the parameters \{$\tau_E, r_E, \tau_I, r_I$\} are drawn from a probability distribution function.
We work with the Gamma distribution for its asymmetric shape, which we skew toward late times:
\begin{equation}
    f(\tau | \alpha,\theta) = \frac{1}{\Gamma(\alpha)\theta^\alpha} \tau^{\alpha - 1} e^{-\tau/\theta} \hspace{2em} \text{       for } \tau > 0
\end{equation}
The shape $\alpha$ and scale $\theta$ respectively control the location of the maximum and sharpness of the Gamma distribution.
Multimodal Gamma distributions admit the interpretation of sub-populations in each state differing in their responses to new information.

\subsection{Agent motion}
\label{sec:agent_motion}
The agents undergo drift and diffusion (Fig. \ref{fig:advectiondiffusion}c) such that the position $\boldsymbol{X}_i(t) \in \Omega$ of agent $i$ evolves according to a continuous-time stochastic process:
\begin{equation}
    \label{eq:cont-time-stoc}
    \mathrm{d}\boldsymbol{X}_i(t)
    = -\nabla_{\boldsymbol{X}_i}\psi_i\,\mathrm{d}t
      + \sqrt{2D}\,\mathrm{d}\boldsymbol{W}_i(t),
\end{equation}
where the potential gradients give rise to deterministic driving forces and the stochastic term models random walks.
In general, the potential experienced by agent $i$ depends on its own position and on the current configuration of the rest of the population, $\mathcal{X}_i = \{\boldsymbol{X}_1,\ldots,\boldsymbol{X}_N\}\backslash\{\boldsymbol{X}_i\}$:
\begin{equation}
    \label{eq:agent_potential_general}
    \psi_i \equiv \psi\big(\boldsymbol{X}_i;\,\mathcal{X}_i\big).
\end{equation}
We denote the relative-position vector and the separation between agents $i$ and $j$ by
\begin{equation}
    \label{eq:relative_position}
    \boldsymbol{r}_{ij} = \boldsymbol{X}_j - \boldsymbol{X}_i,
    \qquad
    r_{ij} = \|\boldsymbol{r}_{ij}\|.
\end{equation}
In this work, $\psi_i$ is the sum of a state-dependent background potential, which depends on the position of agent $i$ alone, and pairwise, isotropic inter-agent potentials:
\begin{equation}
    \label{eq:psi_tot}
    \psi_i
    =
    \underbrace{\Xi_i\,\psi^{\text{BG}}(\boldsymbol{X}_i)}_{\psi_i^{\text{BG}}}
    +
    \underbrace{\sum_{j \neq i,\, r_{ij} \leq r_c} \phi(r_{ij})}_{\psi_i^{\text{A}}},
\end{equation}
where $\psi^{\text{BG}}$ is the background environmental potential (Sec.~\ref{sec:bg-pot}), $\phi(r)$ is a pair potential common to all agents (Sec.~\ref{sec:inter-agent-pot}), and $r_c$ is a cutoff radius.
The indicator $\Xi_i = 1$ if agent $i$ is currently informed and $\Xi_i = 0$ otherwise, so that only informed agents respond to the background potential.
The inter-agent potential is active for all agents regardless of state.
The potential $\psi_i$ has no explicit time dependence; it changes in time only through the agent positions and the information state of agent $i$.
Since $\phi$ depends only on the separation $r_{ij}$, the inter-agent contribution is invariant under translations and rotations of the whole configuration, and the sum over $j$ makes it independent of the ordering of the neighbors.
The reduction of Eq.~\ref{eq:agent_potential_general} to this form is given in Appendix~\ref{app:drift_derivation}.

Taking the gradient of each potential term provides the agent-to-agent and background potential forces:
\begin{equation}
\boldsymbol{F}^\text{A}_i = -\nabla_{\boldsymbol{X}_i}\psi_i^\text{A} \qquad
\boldsymbol{F}^\text{BG}_i = -\nabla_{\boldsymbol{X}_i}\psi_i^\text{BG}
\end{equation}
which, when substituted into Eq.~\ref{eq:cont-time-stoc}, gives:
\begin{equation}
    \label{eq:cont-time-stoc-2}
    \mathrm{d}\boldsymbol{X}_i(t)
    = \left( \boldsymbol{F}_i^A + \boldsymbol{F}_i^{BG} \right)\mathrm{d}t
      + \sqrt{2D}\,\mathrm{d}\boldsymbol{W}_i(t),
\end{equation}

We discuss each term in detail in Sec.~\ref{sec:inter-agent-pot} and Sec.~\ref{sec:bg-pot}.

The last term on the right-hand side of Eq.~\ref{eq:cont-time-stoc-2} scales a standard two-dimensional Wiener process $\boldsymbol{W}_i(t)$, independently for each agent. The random increment $\mathrm{d}\boldsymbol{W}_i(t)$ is drawn from a normal distribution with zero mean and variance $\mathrm{d}t$. The coefficient $\sqrt{2D}$ ensures that the variance of the stochastic displacement matches the macroscopic diffusion limit $2D\mathrm{d}t$.

The applied numerical integration of agent trajectories is performed using an Euler-Maruyama discretization with a time step $\Delta t$:
\begin{equation}
\boldsymbol{X}^{k+1}_i
=
\boldsymbol{X}^{k}_i
+
\mathcal{C}_{\Delta_{\max}}
\!\left(\Delta t\,\boldsymbol{v}^{k}_i\right)
+
\sqrt{2D\Delta t}\,\boldsymbol{\eta}^{k}_i,
\qquad
\boldsymbol{\eta}^{k}_i\sim\mathcal{N}(\boldsymbol{0},\boldsymbol{I}),
\end{equation}
Here $\boldsymbol{v}^{k}_i=\boldsymbol{F}^{{\rm A},k}_i+\boldsymbol{F}^{{\rm BG},k}_i$ is the total deterministic drift velocity at step $k$, with $\boldsymbol{F}^{{\rm BG},k}_i = -\chi_i^k\,\nabla_{\boldsymbol{X}_i}\psi^{\text{BG}}(\boldsymbol{X}_i^k)$, so that $\boldsymbol{F}^{{\rm BG},k}_i=\boldsymbol{0}$ unless agent $i$ is informed at step $k$. %
The drift, and the exposure test of Algorithm~\ref{alg:state_update}, are both evaluated from the positions and states at step $k$, and state transitions made during the step take effect at step $k+1$.
In the implementation the noise is specified through an amplitude $\sigma_\eta$ multiplying $\sqrt{\Delta t}\,\boldsymbol{\eta}^{k}_i$, so that $D = \sigma_\eta^2/2$.
The operator $\mathcal{C}_{\Delta_{\max}}$ clips the deterministic displacement to a maximum norm $\Delta_{\max}$ to ensure stability during close encounters.

\subsection{The inter-agent potential (\texorpdfstring{$\psi^\text{A}$}{psi A})}
\label{sec:inter-agent-pot}

With radial distance $r$, the short-range pair potential $\phi(r)$ in Eq.~\ref{eq:psi_tot} is of Mie/Lennard-Jones form.
We start with the 12-6 Lennard-Jones potential:

\begin{equation}
    \phi_{\mathrm{LJ}}(r)=4\varepsilon\!\left[\Big(\frac{\sigma}{r}\Big)^{12}-\Big(\frac{\sigma}{r}\Big)^{6}\right]
\end{equation}
which generalizes to a two-term Mie potential:
\begin{equation}
    \phi(r)=\varepsilon\!\left[\Big(\frac{\sigma_{\rm rep}}{r}\Big)^{n}-\Big(\frac{\sigma_{\rm attr}}{r}\Big)^{m}\right]
\end{equation}
where $n$ and $m$ are the repulsive and attractive exponents, $\varepsilon$ sets the interaction strength, and $\sigma_{\rm rep}$ and $\sigma_{\rm attr}$ are characteristic length scales (these length scales are set equal in the conventional LJ form).
The conventional LJ prefactor of 4 is the standard normalization when adopting the 12-6 form.
The radial force follows from the negative derivative,
\begin{equation}
    F_r(r)=-\frac{\mathrm d}{\mathrm d r}\phi(r) =\varepsilon\!\left[n\frac{\sigma_{\rm rep}^n}{r^{\,n+1}}-m\frac{\sigma_{\rm attr}^m}{r^{\,m+1}}\right],
\end{equation}
which is positive (repulsive) at short range.
With $\boldsymbol r_{ij}$ and $r_{ij}$ as in Eq.~\ref{eq:relative_position}, the force exerted on agent $i$ by agent $j$ acts along the line joining them,
\begin{equation}
    \label{eq:pairwise_force}
    \boldsymbol F_{ij}
    = -\nabla_{\boldsymbol{X}_i}\phi(r_{ij})
    = -\phi'(r_{ij})\,\frac{\boldsymbol{X}_i-\boldsymbol{X}_j}{r_{ij}}
    = -\varepsilon\!\left[n\frac{\sigma_{\rm rep}^n}{r_{ij}^{\,n+2}}-m\frac{\sigma_{\rm attr}^m}{r_{ij}^{\,m+2}}\right]\boldsymbol r_{ij},
\end{equation}
so that a repulsive interaction ($F_r > 0$) pushes agent $i$ away from agent $j$.

Since $\psi_i^{\text{A}}$ sums the pair potentials of all neighbors within the cutoff radius $r_c$ (Eq.~\ref{eq:psi_tot}), the total deterministic inter-agent driving force on agent $i$ is the corresponding sum of pairwise forces:
\begin{equation}
\boldsymbol{F}_i^{\text{A}} = -\nabla_{\boldsymbol{X}_i} \psi_i^{\text{A}} = \sum_{j \neq i, \, r_{ij} \leq r_c} \boldsymbol{F}_{ij}
\end{equation}

As explicitly shown in this summation, the inter-agent deterministic drift is evaluated strictly over neighbors within the distance $r_c$.
This finite cutoff is standard in molecular and agent-based simulations: combined with neighbor lists it avoids an all-pairs force evaluation, and it removes negligible long-range contributions. We employ an  all-pairs neighbor search (Sec.~\ref{sec:implementation}) and a sharp cutoff.
Interaction forces are abruptly set to zero for $r_{ij} > r_c$ without the use of potential shifting or switching functions.
Consequently, the numerical model should be interpreted as a numerically stabilized, cutoff approximation to the continuous drift dynamics.
When combined with the zero-distance guard and the deterministic step-size ceiling ($\mathcal{C}_{\Delta_{\max}}$ introduced in Section~\ref{sec:agent_motion}), this sharp cutoff improves global simulation stability while only modifying the very short-time motion and effective mobility of agents during close encounters.

\subsection{The background environmental potential field (\texorpdfstring{$\psi^{\text{BG}}$}{psi BG})}
\label{sec:bg-pot}

The background potential force required in Eq.~\ref{eq:cont-time-stoc-2} is the negative spatial gradient of the background environmental potential $\psi^{\text{BG}}$:
\begin{equation}
\boldsymbol{F}_i^{\text{BG}} = -\Xi_i\,\nabla_{\boldsymbol{X}_i} \psi^{\text{BG}}(\boldsymbol{X}_i)
\end{equation}

To construct the background environmental potential $\psi^{\text{BG}}$, we use linear superposition of continuously differentiable analytical potentials, $\psi(\boldsymbol{X})$. By equating the driving force $\boldsymbol{F}_{\text{drift}} = -\nabla_{\boldsymbol{X}}\psi(\boldsymbol{X})$ with the drift velocity through a unit mobility, we already have invoked the principle of gradient flow. It follows that potential maxima are local regions from which agents will be repelled, minima are attractors, and valleys with sloping floors are conduits for agents' downhill flows. Agents congregate at stable extrema where $\nabla_{\boldsymbol{X}}\psi = 0$ and the Hessian matrix is positive-definite ($\nabla_{\boldsymbol{X}}^2\psi \succ 0$), while unstable extrema characterized by a negative-definite Hessian ($\nabla_{\boldsymbol{X}}^2\psi \prec 0$) represent hazards.

For clarity of construction of such potential landscapes we separate the nodal points of repulsion and attraction from the (generally curvilinear and sloping) conduits.

\begin{equation}
\psi^{\text{BG}}(\boldsymbol{X}_i) = \psi_{\text{nodes}}(\boldsymbol{X}_i) + \psi_{\text{conduits}}(\boldsymbol{X}_i)
\end{equation}

\subsubsection{Nodal landscape features}
\label{sec:potentiallandscape}
Maxima and minima are obtained by superposing multi-centered Gaussian functions. To obtain an arbitrary number of maxima and minima, the global potential is endowed with the following terms:

\begin{equation}
    \begin{aligned}
    \psi_{\text{nodes}}(\boldsymbol{X}) ={}&
    \sum_{j} a_{\text{max},j} \exp\left( -\frac{\|\boldsymbol{X}-\boldsymbol{X}_{\text{max},j}\|^2}{2\sigma_{\text{max},j}^2} \right) \\
    &+ \sum_{k} a_{\text{min},k} \exp\left( -\frac{\|\boldsymbol{X}-\boldsymbol{X}_{\text{min},k}\|^2}{2\sigma_{\text{min},k}^2} \right)
    \label{eq:global_landscape_potential}
\end{aligned}
\end{equation}
where $a_{\text{max},j} > 0$ scales the $j$-th maximum centered at $\boldsymbol{X}_{\text{max},j}$, and $a_{\text{min},k} < 0$ dictates the depth of the $k$-th local minimum  at $\boldsymbol{X}_{\text{min},k}$, with widths controlled by the respective variance parameters $\sigma_{\text{max},j}$ and $\sigma_{\text{min},k}$.

\subsubsection{Conduit features}
\label{sec:conduitlandscape}

To arrive at conduits with complex shapes, we augment the environment with low-energy troughs swept along a parameterized curvilinear centerline, denoted as $\mathcal{C}$.
Consider an arbitrary point $\boldsymbol{X}$ near this conduit.
Let $\boldsymbol{q}$ represent the orthogonal projection of $\boldsymbol{X}$ onto the centerline, defined by $\boldsymbol{q} = \operatorname*{argmin}_{\boldsymbol{p} \in \mathcal{C}} \|\boldsymbol{X} - \boldsymbol{p}\|$.
The geometry of the pathway at $\boldsymbol{q}$ establishes a moving local coordinate frame defined by the tangent vector $\boldsymbol{T}$, which dictates the local ``forward'' direction of travel, and the unit normal $\boldsymbol{N}$, pointing toward the inside of the bend.
The spatial displacement between $\boldsymbol{X}$ and the centerline path is given by the vector $\boldsymbol{\delta} = \boldsymbol{X} - \boldsymbol{q} = \ell\,\boldsymbol{N}$, where $\ell$ is the signed lateral distance from the centerline, positive toward the inside of the bend, and the arc-length position along the centerline is given by $s(\boldsymbol{X})$.

A natural starting point is to construct a unified, conservative potential $\psi_{\text{conduit}}$ that simultaneously confines agents to the path and slopes downward along it.
This can be structured as an inverted Gaussian profile parameterized by a geometric scaling factor $B = 1/(2\sigma_c^2)$, a base depth $A$, and a constant slope $g$:

\begin{equation}
    \psi_{\text{conduit}}(\boldsymbol{X}) = - \big(A + g \, s(\boldsymbol{X})\big) \exp\left( -B\Vert\boldsymbol{\delta}\Vert^2 \right)
\end{equation}

Taking the negative spatial gradient $\boldsymbol{F}_{\text{conduit}} = -\nabla \psi_{\text{conduit}}$ yields the corresponding deterministic force.
On a curved centerline, the arc-length gradient is $\nabla s = \boldsymbol{T}/(1-\kappa\ell)$, where $\kappa$ is the local curvature of the centerline (Appendix~\ref{app:conduit_curvature_derivation}), so that

\begin{equation}
    \boldsymbol{F}_{\text{conduit}} = -2B\big(A + g \, s(\boldsymbol{X})\big) \exp\left( -B\Vert\boldsymbol{\delta}\Vert^2 \right) \boldsymbol{\delta}
    +
    \frac{g}{1-\kappa\ell} \exp\left( -B\Vert\boldsymbol{\delta}\Vert^2 \right) \boldsymbol{T}
\end{equation}

We collect the lateral force terms and note that they act to restore agent positions within the conduit, behaving like a transverse spring and confining random walking agents within the valley width:

\begin{equation}
    \boldsymbol{F}_{\text{restore}} =
    -2B\big(A + g \, s(\boldsymbol{X})\big) \exp\left( -B\Vert\boldsymbol{\delta}\Vert^2 \right) \boldsymbol{\delta}
    \label{eq:channel_restoring_force}
\end{equation}

The remaining tangential term acts as a forward drift vector field along the path.
As it is scaled by the same Gaussian envelope, this propulsion is strongest at the valley floor and decays up the valley walls, while propelling the agents along $\boldsymbol{T}$.

\begin{equation}
    \boldsymbol{F}_{\text{forward}} = \frac{g}{1-\kappa\ell} \exp\left( -B\|\boldsymbol{\delta}\|^2 \right) \boldsymbol{T}
    \label{eq:channel_forward_force}
\end{equation}
The factor $1/(1-\kappa\ell)$ strengthens the forward drift on the inside of a bend and weakens it on the outside; it reduces to unity for a straight conduit ($\kappa=0$) or on the centerline ($\ell=0$).

The total deterministic kinematic force exerted by the conduit local landscape is the sum of these components:

\begin{equation}
    \boldsymbol{F}_{\text{conduit}} = \boldsymbol{F}_{\text{restore}} + \boldsymbol{F}_{\text{forward}}
    \label{eq:total_channel_force}
\end{equation}

All simulations reported here use this unified potential; an alternative forcing that removes the growth of lateral stiffness along the conduit is described in Appendix~\ref{app:conduit_alt}.

\section{Results}
\label{sec:appl_numer}

We now study the spatiotemporal evacuation dynamics over complex environmental landscapes at agent scale resolution.
In order to drive the agent dynamics that would arise during an evacuation, we model safe and hazardous regions using the landscape of $\psi^\text{BG}$: safe zones as local potential minima, hazards as local maxima and safe conduit paths as sloping valleys.
In such a form, $\psi^\text{BG}$ represents globally available information.

Our numerical investigations follow a four-tiered progression, systematically isolating the effects of globally available information and inter-agent information exchange on the population's spatiotemporal dynamics.
First, in Sec.~\ref{sec:temporal_heterogeneity}, we establish a baseline for how temporal delays impact the state-transition cascades: $S \to E \to I \to R$.
For this study, the influence of the background potential is silenced by adopting a flat landscape ($\psi^\text{BG}(\boldsymbol{X}) = 0$).
Second, we hold response delays fixed and introduce a $\psi^\text{BG}$ landscape of isolated, isotropic Gaussian wells (hazards and a safe zone), first without conduits (Sec.~\ref{sec:st_dyn_wo_conduits}) and then with curvilinear conduits added as guidance paths (Sec.~\ref{sec:st_dyn_w_conduits}).
Third, in Sec.~\ref{sec:multimodal_study}, we couple the structured environment from the previous study with heterogeneous inter-agent information exchange, governed by multimodal state-transition latency clocks, to observe the emergence of non-equilibrium but frozen dynamics both with and without conduits.
Finally, we conclude with a controlled comparison of landscapes in Sec.~\ref{sec:controls}.
Additional sensitivity studies exploring the effects of agent density, diffusivity, information radius, and boundary conditions are presented in Appendix~\ref{sec:param_sens}.

\subsection{Implementation}
\label{sec:implementation}

We recall that agents  interact through the short-range pair potential $\phi$ of two-term Mie (generalized Lennard-Jones) form, summed over neighbors within a sharp cutoff radius $r_c$ to give $\psi^{\text{A}}$.
For the case studies and the control batch (Table~\ref{tab:case_params}), its strength $\varepsilon = 10^{-6}$ makes $\psi^{\text{A}}$ negligible compared with $\psi^{\text{BG}}$, so agents have no effective cohesion or repulsion.
The delay-study runs of Sec.~\ref{sec:temporal_heterogeneity} used $\varepsilon = 10^{-4}$ and $\sigma = 2^{-1/6}\times10^{-2}$; with $\psi^{\text{BG}} = 0$. Then the only deterministic force is due to the gradient of $\psi^{\text{A}}$.

All simulations in the main text use the domain $\Omega = (0,1)^2$ with periodic boundary conditions and $N = 1000$ agents; the sensitivity studies of Appendix~\ref{sec:param_sens} use the parameters in Table~\ref{tab:lammps_parameters}.
The agent population is initialized with 900 susceptible ($S$) individuals and a seed fraction of 100 informed ($I$) individuals.
In landscape Cases 1--2, exposure delays are drawn from $\tau_E \sim \mathrm{Gamma}(\alpha,\theta), \alpha = 10^{-5}, \theta = 0.75$, whose mean $\alpha\theta = 7.5\times10^{-6}$ is negligible compared with the time step, so that contact proximity triggers effectively instantaneous $S \to E \to I$ transitions.
The remaining baseline parameters are detailed across each case study.
The noise amplitude $\sigma_\eta = 10^{-3}$ corresponds to $D = \sigma_\eta^2 / 2 = 5\times10^{-7}$, so the root-mean-square diffusive displacement over an entire simulation, $\sqrt{4DT} \approx 5\times10^{-3}$ for $T = 15$, is small compared with both the contact radius and the mean agent spacing $\rho^{-1/2} \approx 0.03$.
Motion in these case studies is therefore drift-dominated: for drift speeds $U \approx 0.2$ along the conduits, the P\'eclet number $U r / D$ at the contact scale $r = 0.03$ is of order $10^4$.
Agents that do not respond to $\psi^{\text{BG}}$ (susceptible, exposed and reverted agents) are effectively stationary, and the Lennard-Jones interaction is negligible (Sec.~\ref{sec:inter-agent-pot}).

The implementation neglects the curvature factor $1/(1-\kappa\ell)$ in Eq.~\ref{eq:channel_forward_force}, applying the forward force $g\exp(-B\|\boldsymbol{\delta}\|^2)\,\boldsymbol{T}$, which is exact only for a straight conduit.
For the conduits used here, the resulting error is about 15\% of the peak forward force $g$ (Appendix~\ref{app:conduit_curvature_error}).

Additional computational details are included in Appendix~\ref{app:sim_params}, along with a table that summarizes the parameters of every simulation reported in the main text (Table~\ref{tab:case_params}).

\subsection{Heterogeneity of temporal response and spheres of interaction }
\label{sec:temporal_heterogeneity}

Full run parameters for this section are given in Table~\ref{tab:case_params} (Appendix~\ref{app:sim_params}).
Here, the background potential is switched off ($\psi^{\text{BG}} = 0$), and agents can exhibit  heterogeneity in the rate of absorption of information.
We therefore have probabilistic transition delays governed by Gamma distributions, chosen due to their asymmetric form allowing for weighting toward longer delays before transmissions (see Appendix~\ref{sec:pdf} for example plots).
By thus manipulating the latency of exposure processing ($\tau_E$) and information retention ($\tau_I$), we can obtain distinct state-transition profiles---ranging from rapid-transitioning subpopulations through asymptotically extended latencies---and observe how this temporal heterogeneity alters the population-scale waves of information propagation.
Note that these delay-study runs used a stronger Lennard-Jones interaction ($\varepsilon = 10^{-4}$, a hundredfold larger than the $10^{-6}$ used in later case studies and controls) and a smaller exposure radius ($r_E = 0.03$ versus the effective $0.1$ in Cases 1--2).

We first evaluate the impact of unimodal and bimodal distributions for $\tau_E$ and $\tau_I$.
For this baseline set of simulations, we initialize a population of $N=1000$ agents, each susceptible with probability $0.9$ and informed with probability $0.1$ (about 900 and 100 agents, respectively).
The baseline homogeneous response is modeled using a unimodal Gamma distribution ($\alpha=1.5, \theta=0.75$), with mean transition delay $\alpha\theta \approx 1.1$.
To introduce heterogeneity, we implement a bimodal mixture where 700 agents follow the rapid baseline distribution, and the remaining 300 are drawn from a delayed distribution ($\alpha=4.5, \theta=0.75$) with mean transition delay $\alpha\theta \approx 3.4$.

\begin{figure}[tbp]
    \centering
    \includegraphics[width=\linewidth]{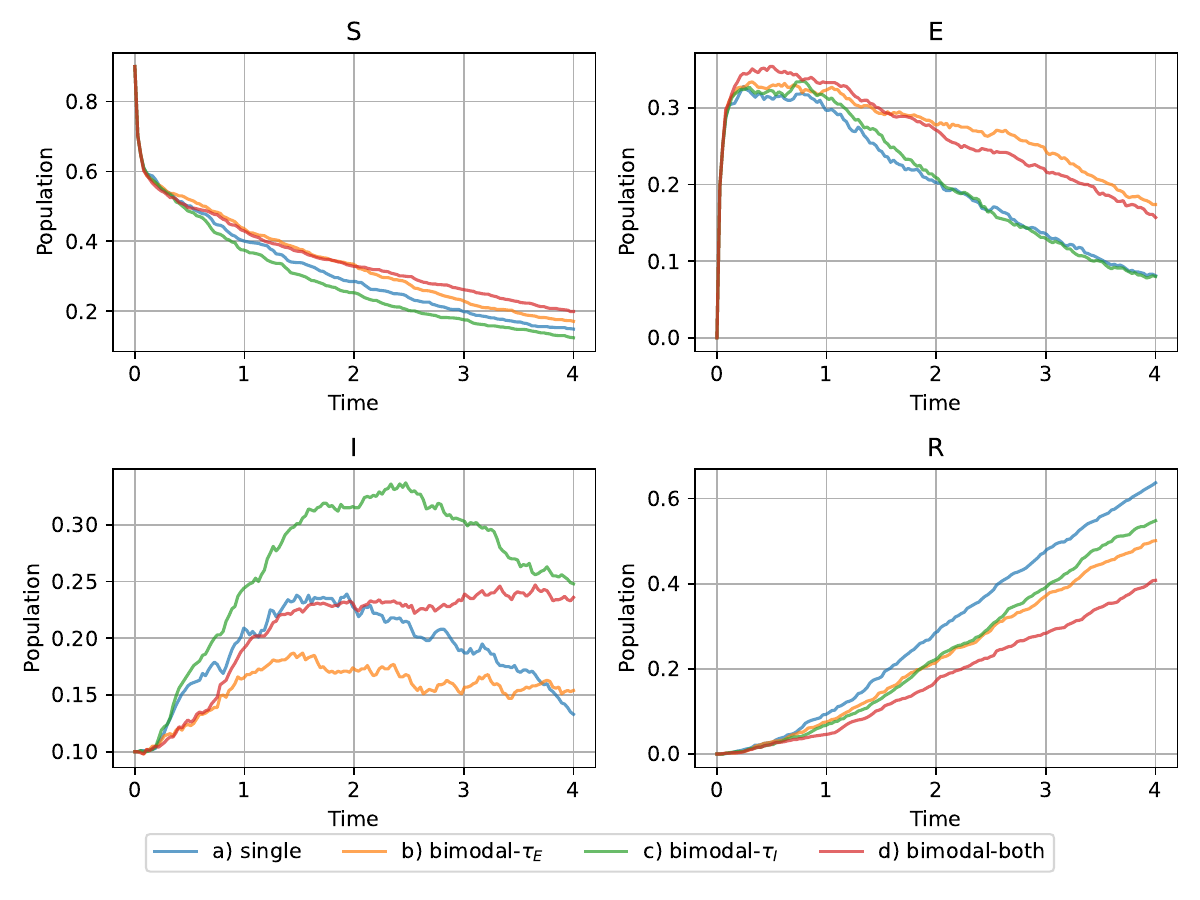}
    \caption{Effect of bimodal distributions for $\tau_E$ (exposed to informed delay) and $\tau_I$ (informed to reverted delay) on population-level response. Single: $\mathrm{Gamma}(\alpha,\theta), \alpha = 1.5, \theta = 0.75$. Bimodal: 700 agents drawn from $\mathrm{Gamma}(\alpha,\theta), \alpha = 1.5, \theta = 0.75$ and 300 from $\mathrm{Gamma}(\alpha,\theta), \alpha = 4.5, \theta = 0.75$. Evaluated with 1000 agents over 150 timesteps, with exposure radius $r_E = 0.03$ and information radius $r_I = 0$.}
    \label{fig:bimodal_tau_i}
\end{figure}

This transition delay critically alters the availability of agents within the transmission cascade.
When a subset exhibits delayed information acceptance (a bimodal $\tau_E$), a larger portion of the population becomes trapped in the exposed state.
This temporary bottleneck prevents them from actively broadcasting to other agents, significantly reducing the peak of the informed population.
Conversely, prolonged retention (a bimodal $\tau_I$) mitigates this reduction; by extending the time agents spend in the informed state before reverting, their density rises and remains high  longer to deplete the susceptible pool.
Figure \ref{fig:bimodal_tau_i} illustrates the  multifactorial dynamics that emerge when both transitions ($\tau_E$ and $\tau_I$) are bimodal.
While the overall supply of informed agents is suppressed by the initial exposure delay, the prolonged retention partially offsets this, creating a long-lasting plateau in their density.

\begin{figure}[tbp]
    \centering
    \includegraphics[width=1.0\linewidth]{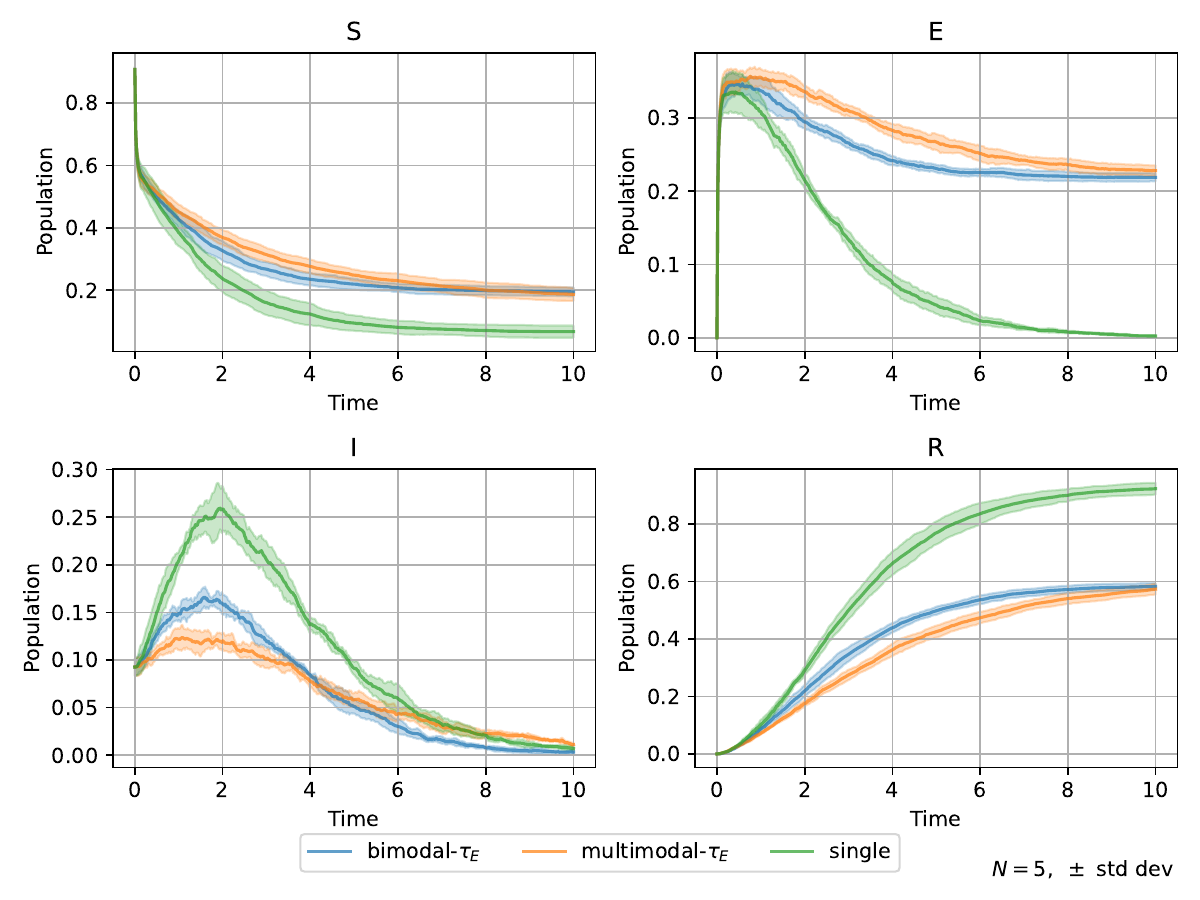}
    \caption{Effect of unimodal, bimodal, and multimodal distributions for $\tau_E$. Five repeated runs, error bars show $\pm$ 1 standard deviation. Single: $\tau_E \sim \mathrm{Gamma}(\alpha,\theta), \alpha = 1.5, \theta = 0.75$. Bimodal: 700 agents from $\mathrm{Gamma}(\alpha,\theta), \alpha = 1.5, \theta = 0.75$ and 300 from $\mathrm{Gamma}(\alpha,\theta), \alpha = 1000, \theta = 0.75$. Multimodal: 500 agents from $\mathrm{Gamma}(\alpha,\theta), \alpha = 1.5, \theta = 0.75$, 200 from $\mathrm{Gamma}(\alpha,\theta), \alpha = 4.5, \theta = 0.75$ and 300 from $\mathrm{Gamma}(\alpha,\theta), \alpha = 1000, \theta = 0.75$. Evaluated with 1000 agents over 500 timesteps ($\Delta t = 0.02$), with exposure radius $r_E = 0.03$ and information radius $r_I = 0$.} %
    \label{fig:multimodal_tau_e}
\end{figure}

Building upon this, we explore an extreme scenario where a significant subset of the population (300 agents) exhibits infinite latency and effectively never transitions.
To model this, we utilize a multimodal distribution (Figure \ref{fig:multimodal_tau_e}) that introduces a third, highly latent peak ($\alpha = 1000$) in the $\tau_E$ distribution.
In this scenario, about 190 agents remain susceptible at $t = 10$ and a further 220 remain exposed. The exposed count is close to the latent share ($300$ of $1000$ agents) of the roughly 710 agents that are ever exposed ($0.3 \times 710 \approx 213$).
The final fractions are close to those of the bimodal case in Fig.~\ref{fig:multimodal_tau_e}, which contains the same 300-agent latent component; the exposed plateau is set by this shared latent fraction rather than by a saturation of spatial transmission.
Figure \ref{fig:multimodal_tau_e} also displays the mean and standard deviation of population dynamics over five Monte Carlo runs with random initial conditions drawn from uniform distributions.

\subsection{Spatiotemporal dynamics under a potential without conduits}
\label{sec:st_dyn_wo_conduits}

With the effects of heterogeneity on temporal response established, we proceed to exploring the impacts of background potential on agent dynamics.
Our baseline case study models agent flow through a landscape without conduits, using a homogeneous (unimodal) response-delay profile.
The background environmental potential field consists exclusively of superposed isotropic Gaussian components:

\begin{equation}
    \psi^{\text{BG}}(\boldsymbol{X}) = \sum_{j=1}^{3} a_{\text{max},j} \exp\left( -\frac{\|\boldsymbol{X}-\boldsymbol{X}_{\text{max},j}\|^2}{2\sigma_{\text{max},j}^2} \right) + a_{\text{min}} \exp\left( -\frac{\|\boldsymbol{X}-\boldsymbol{X}_{\text{min}}\|^2}{2\sigma_{\text{min}}^2} \right)
\end{equation}
where three distinct maxima ($a_{\text{max},j} = 0.1, \sigma_{\text{max},j} = 0.1$) are positioned  near the domain boundaries at coordinates $\boldsymbol{X}_{\text{max},1} = \langle 0.1, 0.1 \rangle^\top$, $\boldsymbol{X}_{\text{max},2} = \langle 0.1, 0.9 \rangle^\top$, and $\boldsymbol{X}_{\text{max},3} = \langle 0.9, 0.9 \rangle^\top$. A single global minimum ($a_{\text{min}} = -0.1, \sigma_{\text{min}} = 0.10$) represents a safe zone located at the center, $\boldsymbol{X}_{\text{min}} = \langle 0.5, 0.5 \rangle^\top$.

\begin{figure}[tbp]
    \centering
    \begin{subfigure}{0.48\linewidth}
        \centering
        \includegraphics[width=\linewidth]{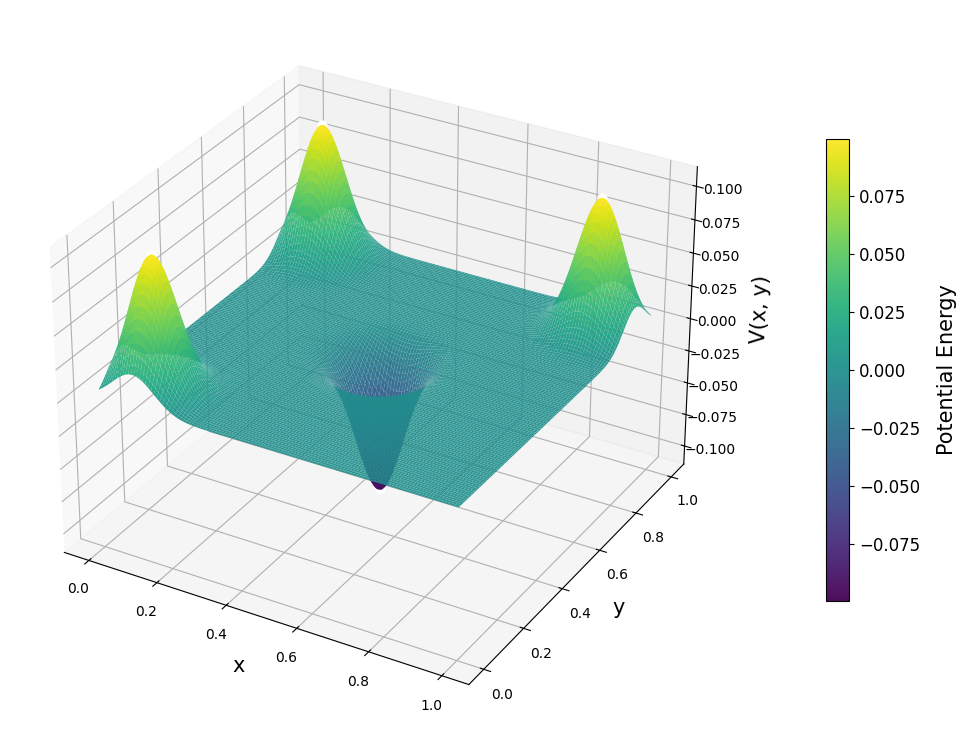}
        \caption{Gaussian-only potential landscape.}
        \label{fig:gaussian_surface_plots}
    \end{subfigure}
    \hfill
    \begin{subfigure}{0.48\linewidth}
        \centering
        \includegraphics[width=\linewidth]{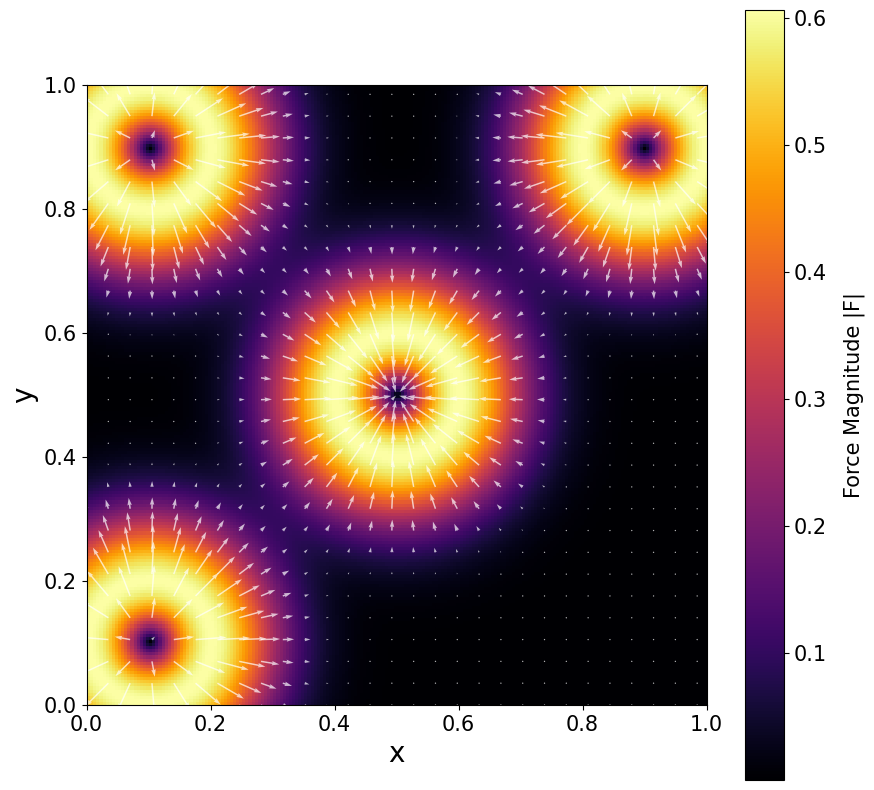}
        \caption{Radial drift vector field.}
        \label{fig:gaussian_drift_fields}
    \end{subfigure}
    \caption{Potential landscape and kinematics of the conduit-free environmental configuration. (a) Potential landscape featuring three local maxima (hazards) and a global minimum (safe zone). (b) The corresponding driving force vectors showing locally symmetric, radial drift velocities  radiating from the hazards and into the safe zone.}
    \label{fig:gaussian_landscape_props}
\end{figure}

The population's initial transition latencies are drawn from a Gamma distribution featuring rapid information processing timescales ($\tau_E \sim \mathrm{Gamma}(\alpha,\theta), \alpha = 10^{-5}, \theta = 0.75$), ensuring swift collective mobilization, while $\tau_I \sim \mathrm{Gamma}(\alpha, \theta), \alpha = 800, \theta = 0.75$, whose mean of 600 greatly exceeds $T = 15$, ensures that no informed agent reverts during the simulation.
The exposure radius $r_E = 0.3$ exceeds the neighbor cutoff $r_c = 0.1$, so the effective contact radius in Cases 1 and 2 is $0.1$ (about 31 agents within range at $\rho = 1000$).
The resulting potential landscape  and its corresponding deterministic drift vector field $\boldsymbol{F}^{\text{BG}} = -\nabla_{\boldsymbol{X}}\psi^\text{BG}(\boldsymbol{X})$ are in Fig.~\ref{fig:gaussian_landscape_props}.

\begin{figure}[tbp]
    \centering
    \begin{subfigure}[t]{0.32\textwidth}
        \centering
        \includegraphics[width=\linewidth]{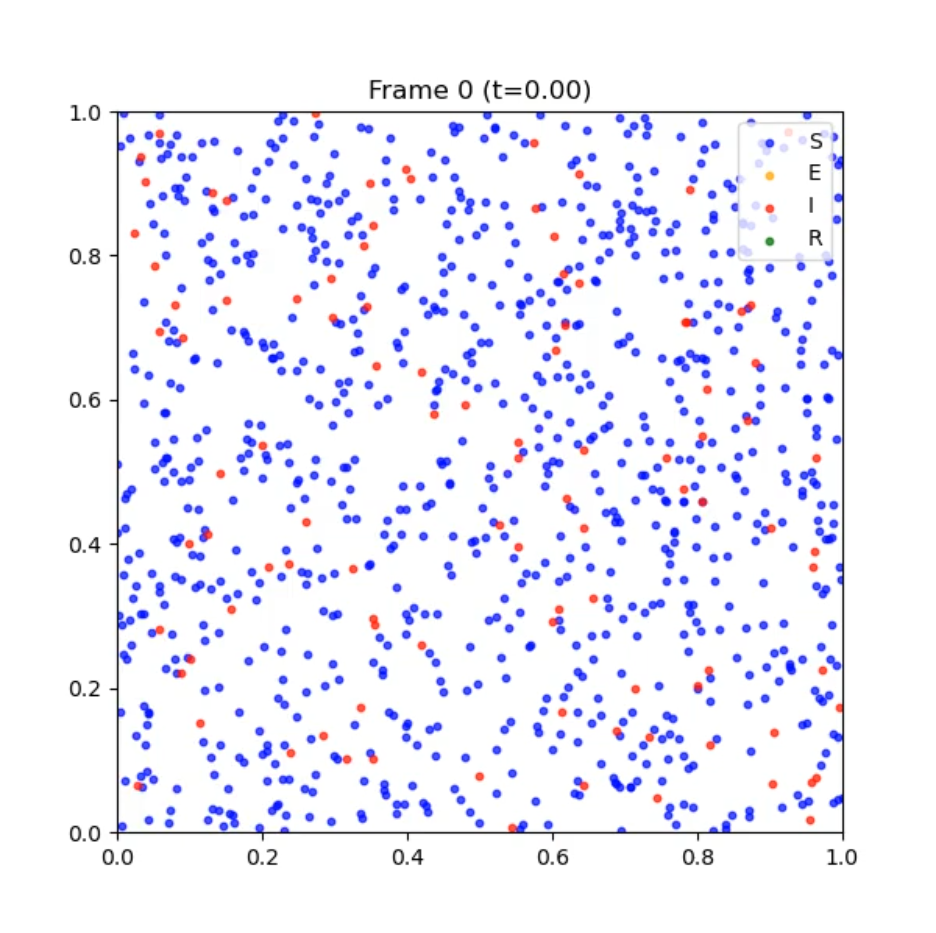}
        \caption{$t=0.00$}
        \label{fig:gauss_snapshot_0}
    \end{subfigure}
    \hfill
    \begin{subfigure}[t]{0.32\textwidth}
        \centering
        \includegraphics[width=\linewidth]{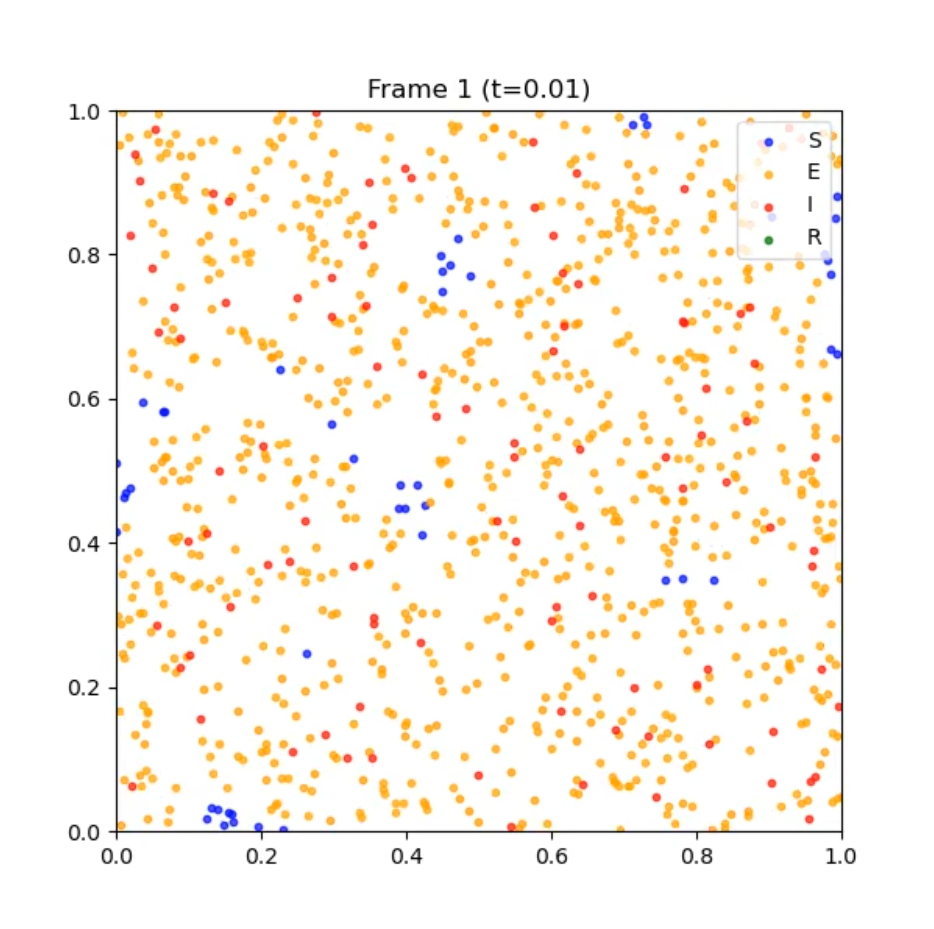}
        \caption{$t=0.01$}
        \label{fig:gauss_snapshot_1}
    \end{subfigure}
    \hfill
    \begin{subfigure}[t]{0.32\textwidth}
        \centering
        \includegraphics[width=\linewidth]{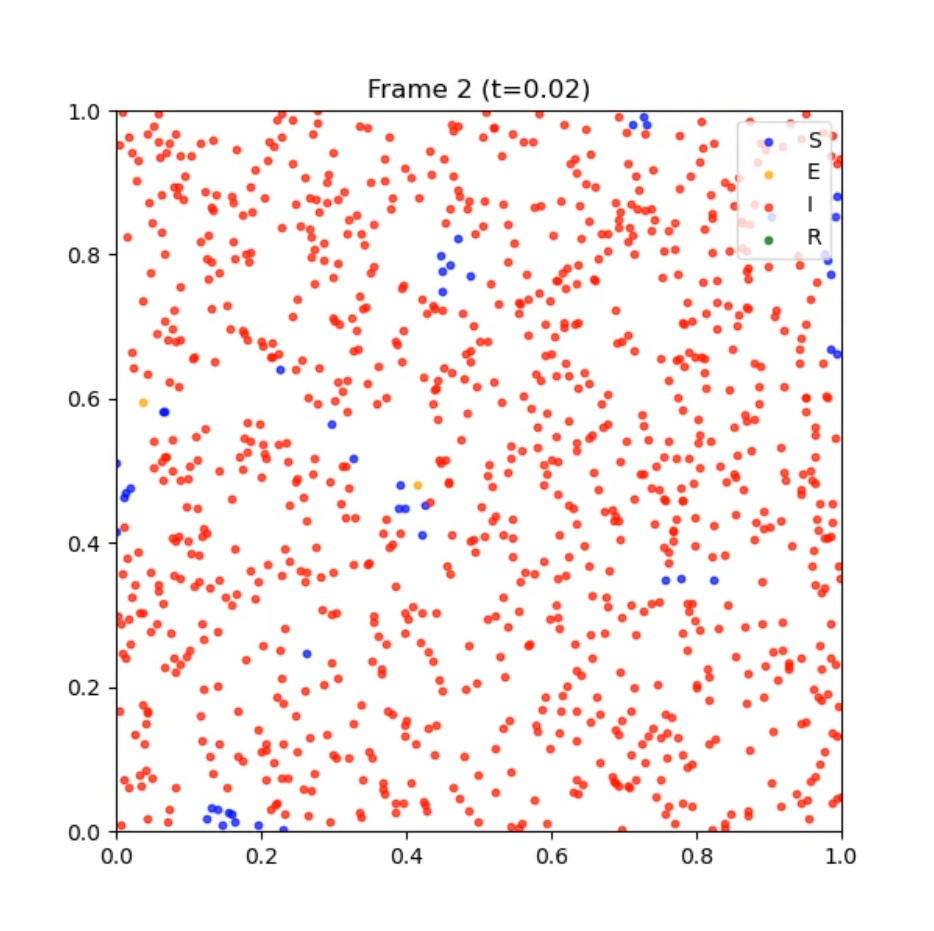}
        \caption{$t=0.02$}
        \label{fig:gauss_snapshot_2}
    \end{subfigure}
    \vspace{0.3cm}
    \begin{subfigure}[t]{0.32\textwidth}
        \centering
        \includegraphics[width=\linewidth]{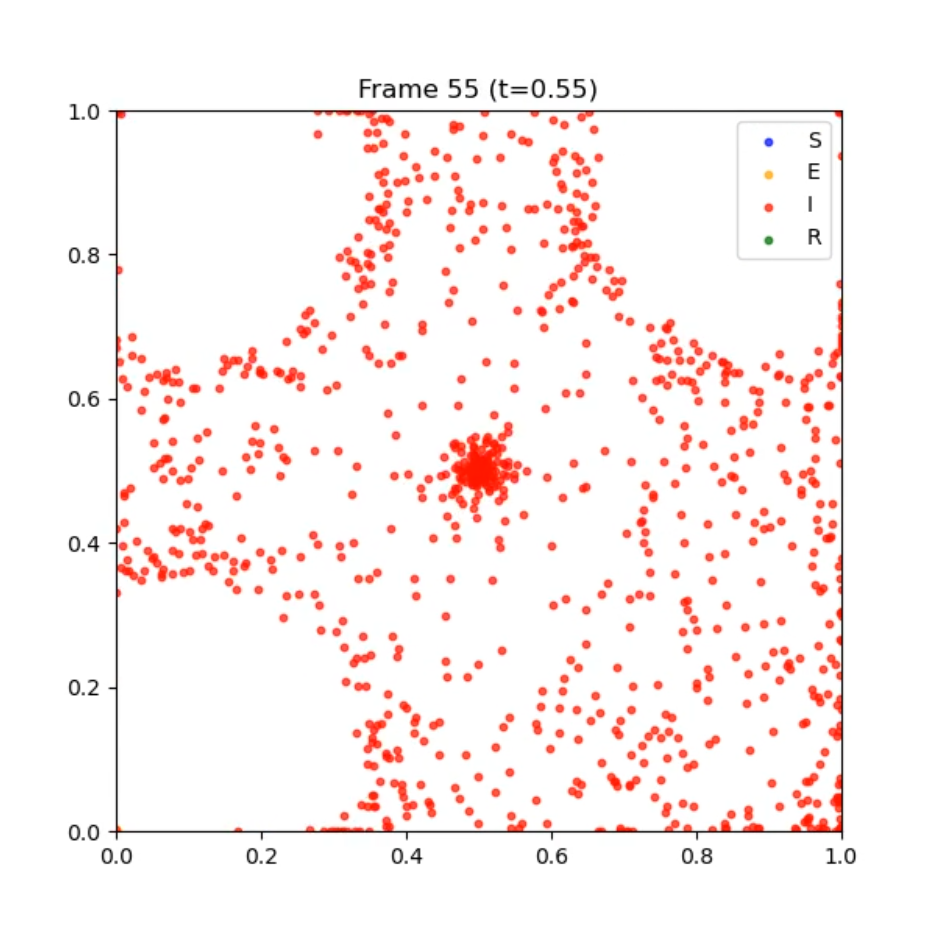}
        \caption{$t=0.55$}
        \label{fig:gauss_snapshot_3}
    \end{subfigure}
    \hfill
    \begin{subfigure}[t]{0.32\textwidth}
        \centering
        \includegraphics[width=\linewidth]{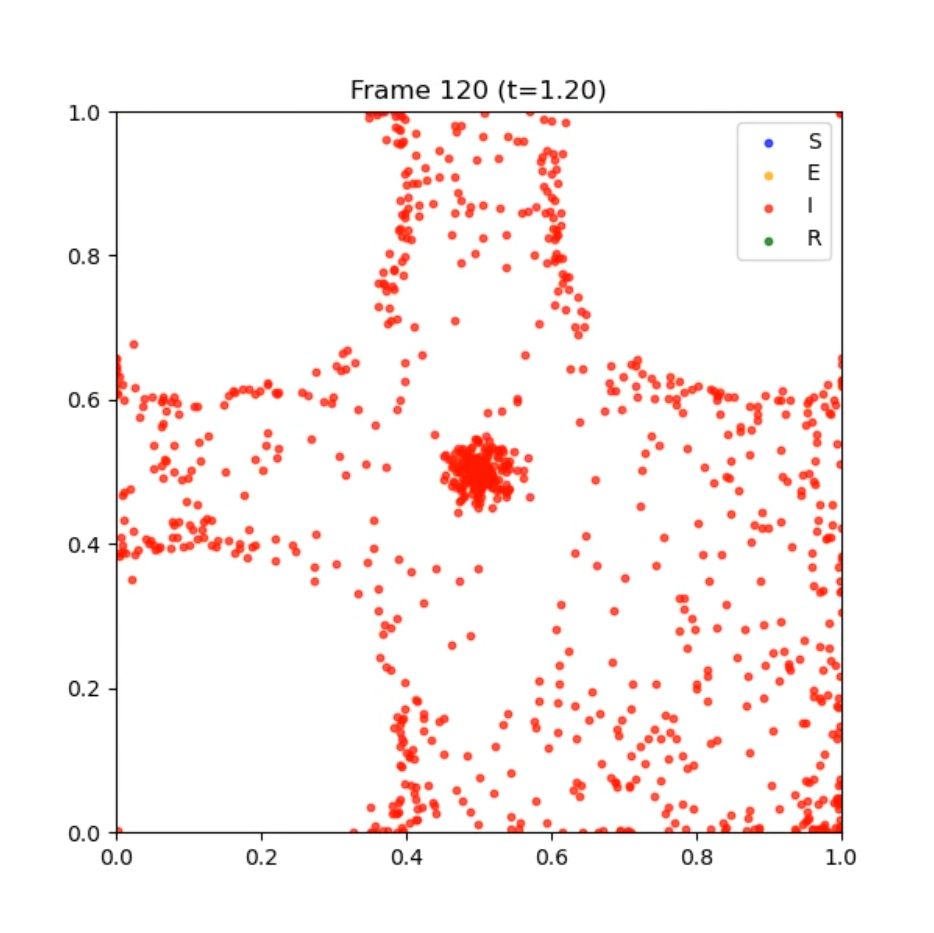}
        \caption{$t=1.20$}
        \label{fig:gauss_snapshot_4}
    \end{subfigure}
    \hfill
    \begin{subfigure}[t]{0.32\textwidth}
        \centering
        \includegraphics[width=\linewidth]{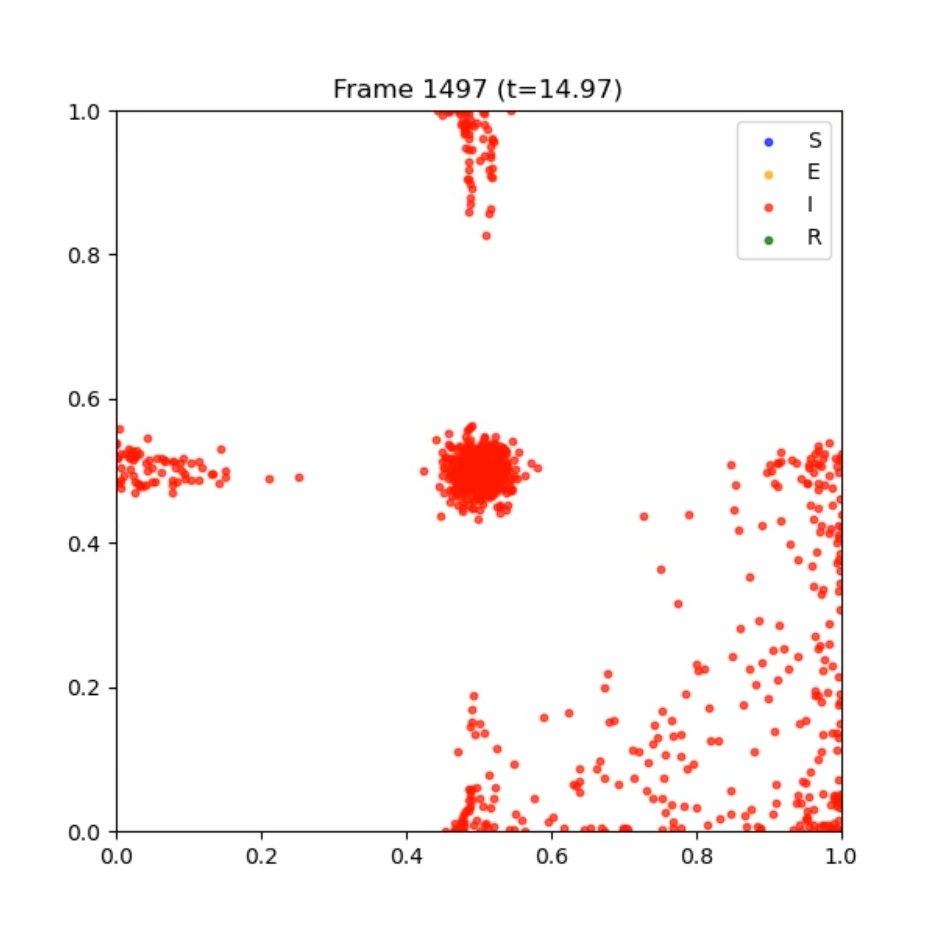}
        \caption{$t=14.97$}
        \label{fig:gauss_snapshot_5}
    \end{subfigure}
    \caption{Time-series snapshots of evacuation dynamics under the potential landscape of Figure \ref{fig:gaussian_landscape_props} with only maxima and minima. Agents rapidly transition from susceptible (blue) to informed (red) under a single Gamma distribution ($\tau_E \sim \mathrm{Gamma}(\alpha,\theta), \alpha = 10^{-5}, \theta = 0.75$; $\tau_I \sim \mathrm{Gamma}(\alpha, \theta), \alpha = 800, \theta = 0.75$, no reversion within the simulation) and exposure/information radii $r_E = 0.3$, $r_I = 0.0$ (effective contact radius $0.1$, set by the neighbor cutoff), respectively. Due to the absence of conduits, agents scatter in a locally radial manner, leading to the observed global pooling patterns. Agents on the plateaus move toward the minimum only under the weak residual gradient of its Gaussian tail.}
    \label{fig:gaussian_sim_snapshots}
\end{figure}

The spatiotemporal evolution of the agent population under this setup appears in Fig.~\ref{fig:gaussian_sim_snapshots}.
Once informed, agents become responsive to the background potential and drift accordingly; inter-agent Lennard-Jones forces remain negligible at the interaction strength used (Sec.~\ref{sec:inter-agent-pot}).

The kinetic trajectories reveal obvious structural limitations of potential fields without conduits.
As the driving forces are locally radial, agents down-gradient of the hazards are expelled outward into the open plains of the domain.
Without guiding conduits, agents disperse widely and form scattered, high-density crowds.

Crucially, agents that wander into the flat potential plateaus---such as the bottom-right quadrant---experience only a small potential gradient.
In that bottom-right quadrant the drift speed is about $0.03$, compared with about $0.6$ near the hazards, and diffusion contributes little at this noise level.
Agents therefore approach the central well only slowly, under the residual gradient of its Gaussian tail.
This results in broad, disorganized and slow pooling of agents in the minimum.

\subsection{Valley-augmented potential landscape}
\label{sec:st_dyn_w_conduits}

Our second case study implements architected conduits while maintaining an identical unimodal population profile ($\tau_E \sim \mathrm{Gamma}(\alpha,\theta), \alpha = 10^{-5}, \theta = 0.75$; $\tau_I \sim \mathrm{Gamma}(\alpha, \theta), \alpha = 800, \theta = 0.75$, no reversion within the simulation).
The background potential is augmented by superposing continuous, curvilinear Gaussian valleys leading from the local maxima to the global minimum at $\boldsymbol{X}_{\text{min}} = \langle 0.5, 0.5 \rangle^\top$.

\begin{figure}[tbp]
    \centering
    \begin{subfigure}{0.48\linewidth}
        \centering
        \includegraphics[width=\linewidth]{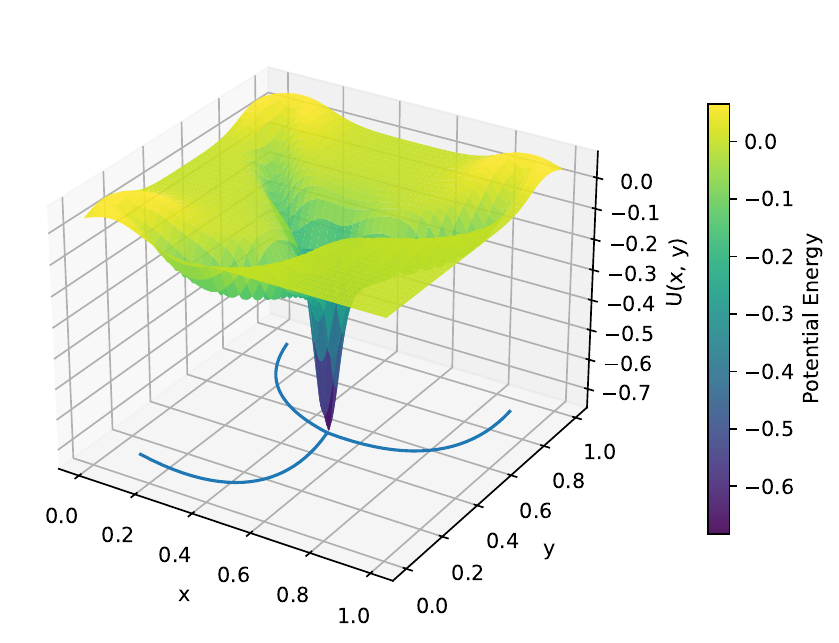}
        \caption{Valley-augmented potential topography.}
        \label{fig:channel_surface_plots}
    \end{subfigure}
    \hfill
    \begin{subfigure}{0.48\linewidth}
        \centering
        \includegraphics[width=\linewidth]{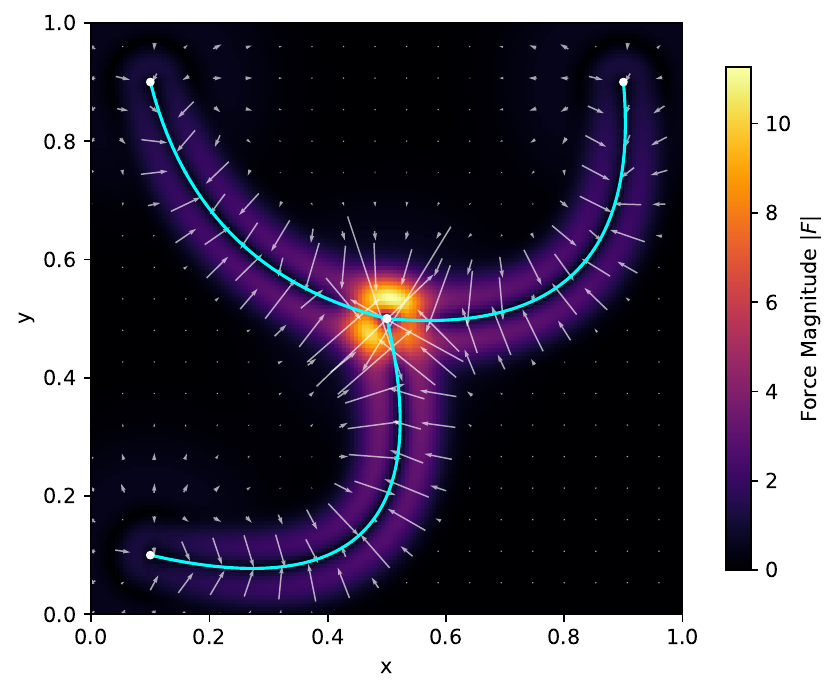}
        \caption{Structured valley vector field.}
        \label{fig:channel_drift_fields}
    \end{subfigure}
    \caption{Potential landscape and force field  of the valley-augmented potential. (a) The  potential  landscape features  valleys with sloping floors connecting maxima  to the central minimum. (b) The drift force field shows highly localized vectors transverse to the valleys  and strong driving forces along the valleys toward the minimum.}
    \label{fig:channel_landscape_props}
\end{figure}

The valleys are obtained by the inverted Gaussian cross-section swept along quadratic Bezier curves.
Far from these valleys, the background potential behaves identically to the preceding section. However, the channel force kinematics are activated within a valley's width: a lateral restoring force $\boldsymbol{F}_{\text{restore}}$ pulls wandering particles onto the centerline track like a transverse spring, while a uniform forward propulsion force $\boldsymbol{F}_{\text{forward}}$ drives them tangentially down the conduit.
The resulting conduit-augmented potential landscape and its highly localized drift vector fields are in Fig.~\ref{fig:channel_landscape_props}.

\begin{figure}[tbp]
    \centering
    \begin{subfigure}[t]{0.32\textwidth}
        \centering
        \includegraphics[width=\linewidth]{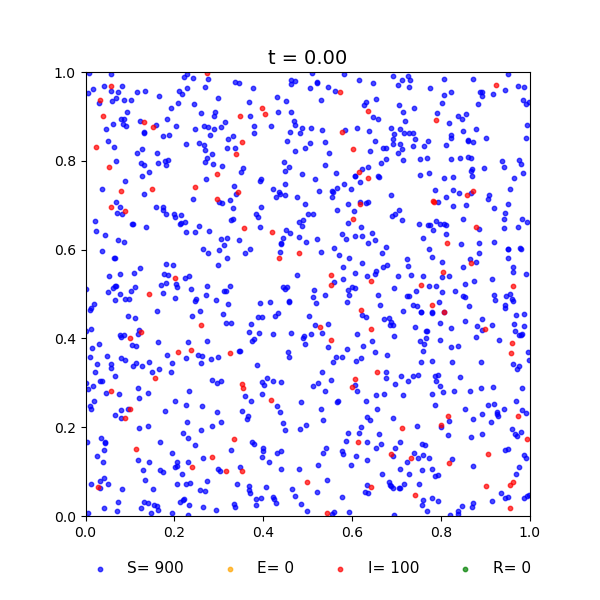}
        \caption{$t=0.00$}
        \label{fig:chan_snapshot_0}
    \end{subfigure}
    \hfill
    \begin{subfigure}[t]{0.32\textwidth}
        \centering
        \includegraphics[width=\linewidth]{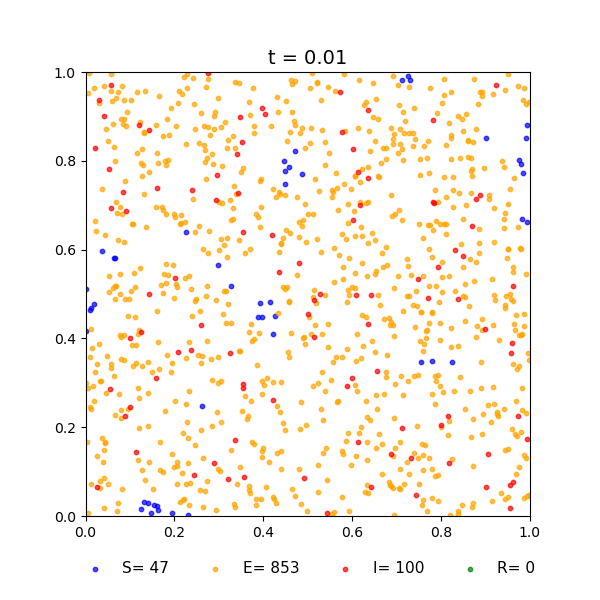}
        \caption{$t=0.01$}
        \label{fig:chan_snapshot_1}
    \end{subfigure}
    \hfill
    \begin{subfigure}[t]{0.32\textwidth}
        \centering
        \includegraphics[width=\linewidth]{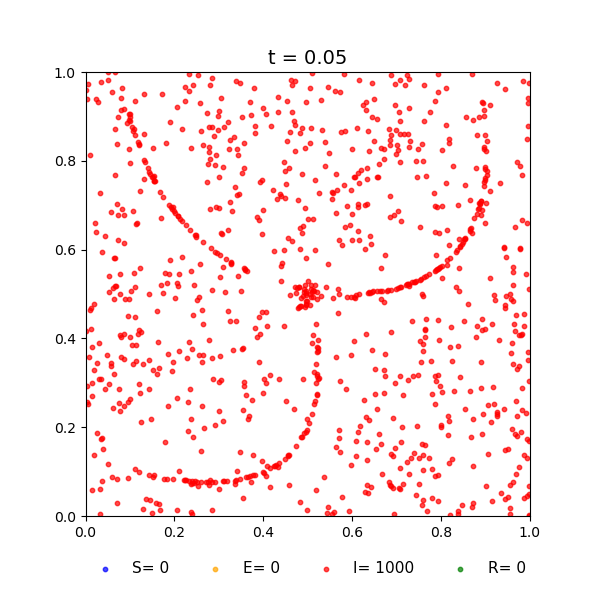}
        \caption{$t=0.05$}
        \label{fig:chan_snapshot_2}
    \end{subfigure}
    \vspace{0.3cm}
    \begin{subfigure}[t]{0.32\textwidth}
        \centering
        \includegraphics[width=\linewidth]{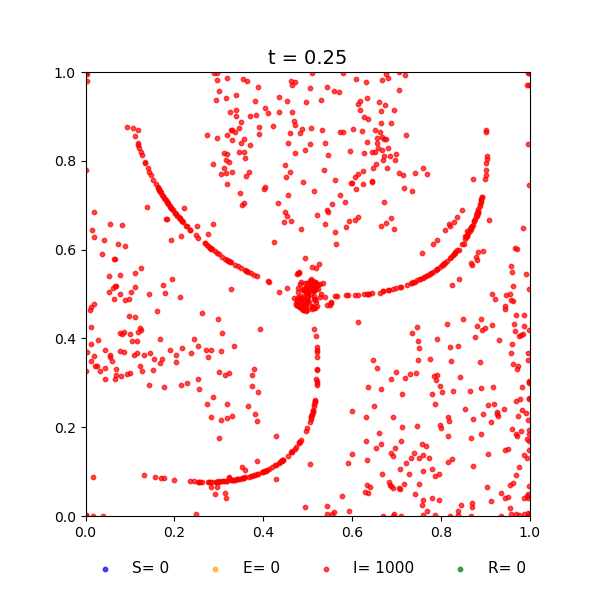}
        \caption{$t=0.25$}
        \label{fig:chan_snapshot_3}
    \end{subfigure}
    \hfill
    \begin{subfigure}[t]{0.32\textwidth}
        \centering
        \includegraphics[width=\linewidth]{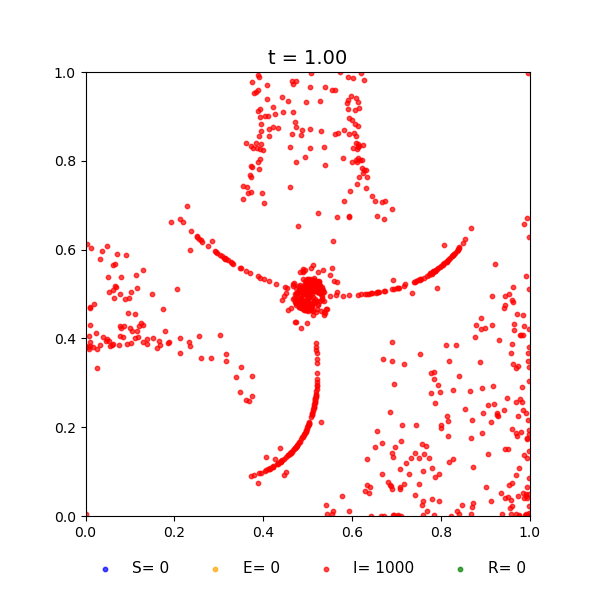}
        \caption{$t=1.00$}
        \label{fig:chan_snapshot_4}
    \end{subfigure}
    \hfill
    \begin{subfigure}[t]{0.32\textwidth}
        \centering
        \includegraphics[width=\linewidth]{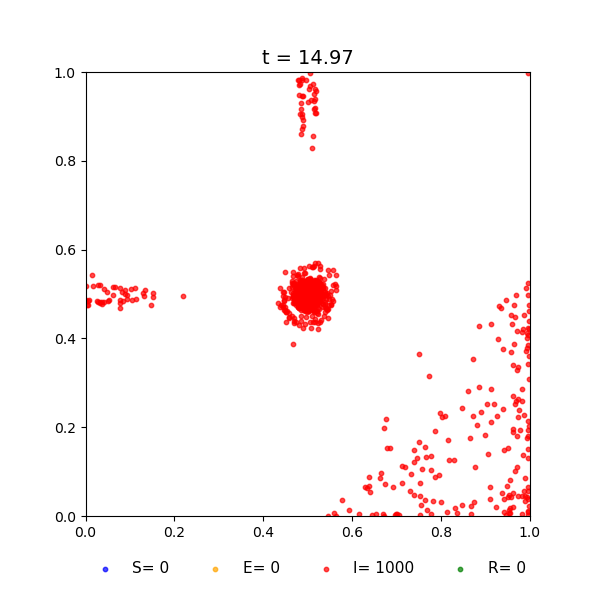}
        \caption{$t=14.97$}
        \label{fig:chan_snapshot_5}
    \end{subfigure}
    \caption{Time-series snapshots of evacuation dynamics within the path-augmented potential landscape.
    All other parameters mimic those shown in the valley-less landscape experiment from Fig \ref{fig:gaussian_sim_snapshots}.
    The valleys replace most of the radial scattering from the maxima with transport along organized conduits and accelerate the population's evacuation; agents that leave the hazards away from the conduit entrances still reach the plateaus (Panel f).} %
    \label{fig:channel_sim_snapshots}
\end{figure}

Figure~\ref{fig:channel_sim_snapshots} shows the resulting flow.
After the same initial state-transition sequence as Case 1, informed agents concentrate within the valleys and are organized into streamlined flows toward the minimum, rather than dispersing radially from the hazards.
The conduits carry most informed agents from the hazards to the central minimum without passing through the low-gradient plateaus.
This organization arises from the potential landscape alone, without discrete routing rules.

\subsection{Multimodal distribution of transition latencies in potential landscapes}
\label{sec:multimodal_study}

Having studied the effect of globally available path information on spatial routing under rapid  transitions between information states, we now revive the influence of distributions in transition latencies.
Our final case study reintroduces the multimodal Gamma distribution from Sec.~\ref{sec:temporal_heterogeneity} to govern the exposure delays $\tau_E$, with $\tau_I \sim \mathrm{Gamma}(\alpha, \theta), \alpha = 14, \theta = 0.75$ (mean 10.5) and contact radii $r_E = 0.03$, $r_I = 0$.
This allows us to model a heterogeneous population having distinct rates of transition through the $S\to E\to I\to R$ cascades navigating the valley-augmented environment potential.

The population's internal clocks are drawn from a multimodal mixture distribution, where 500 agents adopt rapid transitions ($\alpha_1 = 1.5, \theta_1 = 0.75$), 200 delayed transitions ($\alpha_2 = 4.5, \theta_2 = 0.75$), and 300 effectively do not transition within the simulation ($\alpha_3 = 1000, \theta_3 = 0.75$).
This variation introduces a freezing of the dynamics as agents remain trapped in earlier states without transition to being informed.
Furthermore, unlike in Cases 1 and 2, where no agent reverts within the simulation, we see significant transitions from $I$ to $R$ at the end of the simulation.

\begin{figure}[tbp]
    \centering
    \begin{subfigure}[t]{0.32\textwidth}
        \centering
        \includegraphics[width=\linewidth]{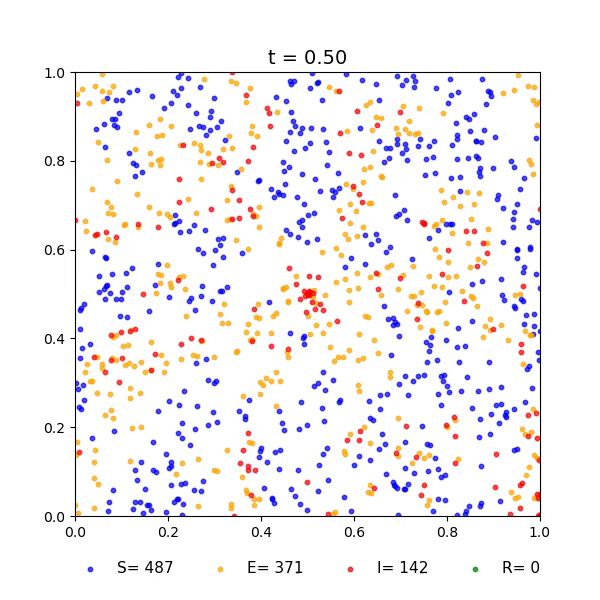}
        \caption{no channel, $t=0.50$}
        \label{fig:multi_snapshot_0}
    \end{subfigure}
    \hfill
    \begin{subfigure}[t]{0.32\textwidth}
        \centering
        \includegraphics[width=\linewidth]{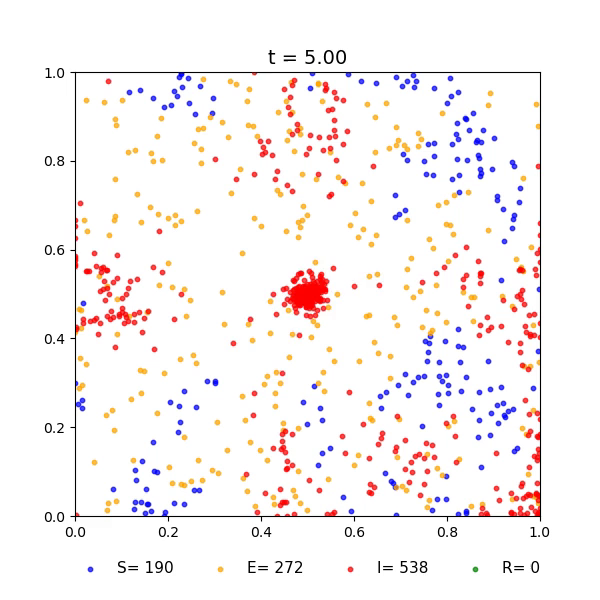}
        \caption{no channel, $t=5.00$}
        \label{fig:multi_snapshot_1}
    \end{subfigure}
    \hfill
    \begin{subfigure}[t]{0.32\textwidth}
        \centering
        \includegraphics[width=\linewidth]{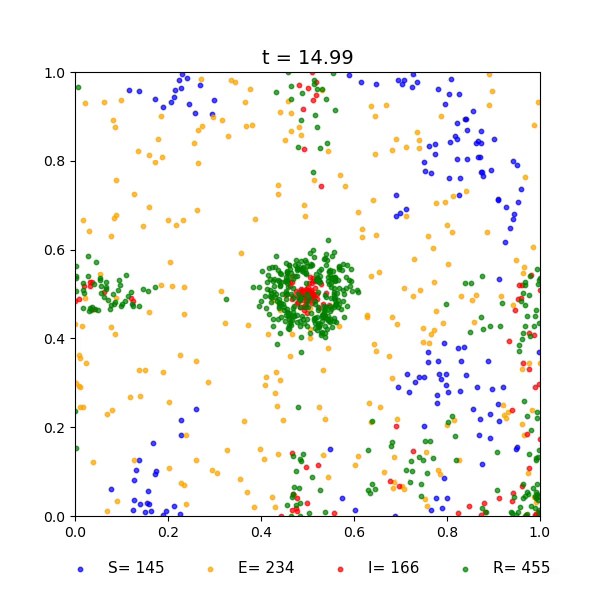}
        \caption{no channel, $t=14.99$}
        \label{fig:multi_snapshot_2}
    \end{subfigure}
    \vspace{0.3cm}
    \begin{subfigure}[t]{0.32\textwidth}
        \centering
        \includegraphics[width=\linewidth]{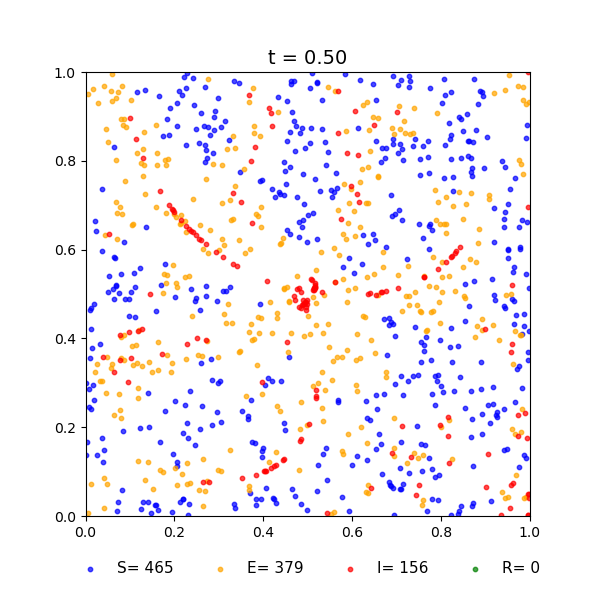}
        \caption{with channel, $t=0.50$}
        \label{fig:multi_snapshot_3}
    \end{subfigure}
    \hfill
    \begin{subfigure}[t]{0.32\textwidth}
        \centering
        \includegraphics[width=\linewidth]{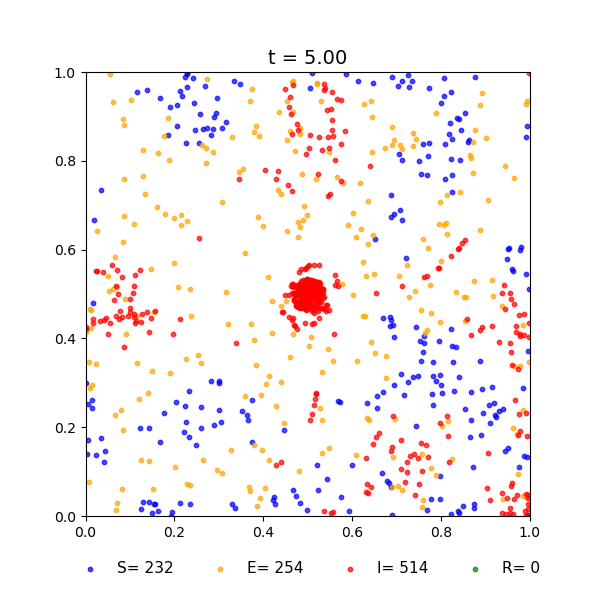}
        \caption{with channel, $t=5.00$}
        \label{fig:multi_snapshot_4}
    \end{subfigure}
    \hfill
    \begin{subfigure}[t]{0.32\textwidth}
        \centering
        \includegraphics[width=\linewidth]{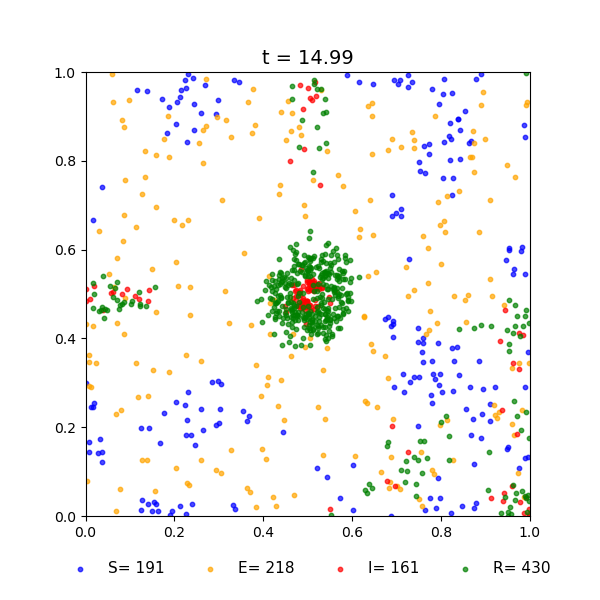}
        \caption{with channel, $t=14.99$}
        \label{fig:multi_snapshot_5}
    \end{subfigure}
    \caption{Time-series snapshots of evacuation dynamics under a multimodal distribution of transition latencies. (a) through (c) illustrate the unarchitected Gaussian landscape, while (d) through (f) include evacuation paths.}
    \label{fig:channel_multimodal_snapshots}
\end{figure}

The macro-scale consequences of this informational-kinetic coupling are visible in the agents' flow (Fig.~\ref{fig:channel_multimodal_snapshots}).
Despite the presence of architected, fast conduits toward the minimum, their utility is lost without the robust supply of a large subpopulation of informed agents to traverse these paths. The global population's dynamics slow down due to delayed transitions.
Because the late-transitioning agents experience an extended latency clock in the exposed state ($E$), they remain unaffected by the background drift field during the early stages of the simulation.
While informed agents (red) immediately consolidate within the valleys and travel swiftly along these conduits to the central target zone at the potential minimum, the slow-responding agents (blue/orange) remain kinetically trapped in the immediate vicinity of the hazards at the local maxima.
These stranded exposed agents remain essentially stationary near the maxima and do not respond to the environmental potential gradients until their delayed internal clocks trigger the transition to the informed state.

This coupling manifests as a multi-phase transport phenomenon: a high-velocity, organized stream of early-transitioning agents alongside a lagging, dispersed tail of exposed agents.
Consequently, the final population distribution has a number of susceptible and exposed agents stagnating on potential plateaus representing regions impervious to globally propagated information.
This demonstrates that the heterogeneity of state transition latencies can control global spatiotemporal dynamics of agent motions independently of the global propagation of information.
It also highlights the importance of modeling the global propagation of information, here via the potential landscape, alongside distribution of state transition latencies as a fully-coupled nonlinear system.

Section~\ref{sec:controls} quantifies this marooning directly by comparing paired runs that differ only in $\psi^{\text{BG}}$.
Conduits raise the safe-zone fraction relative to the unguided landscape, but also leave more agents never exposed.
The apparent marooning highlighted above is therefore not an inherent cost of evacuating, but a comparative effect of the guidance mechanism itself.

\subsection{Controlled comparison of landscapes}
\label{sec:controls}

\begin{figure}[tbp]
    \centering
    \includegraphics[width=1.0\linewidth]{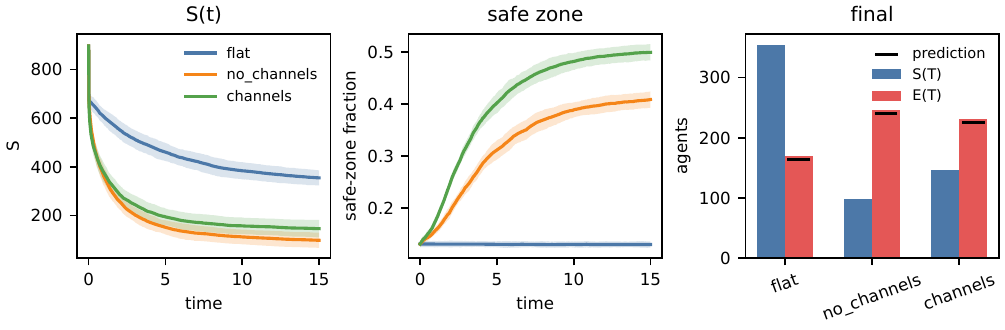}
    \caption{Control summary. (a) $s(t)$ per landscape, seed mean $\pm$ 1 s.d.; (b) safe-zone fraction vs time; (c) final $S$ and $E$ with the latent-fraction prediction.}
    \label{fig:control_summary}
\end{figure}

We conclude our results by isolating the effect of guidance with a controlled comparison.
We begin with the multimodal delay profile and hazard geometry of Case 3 (Sec.~\ref{sec:multimodal_study}) and run it on three background landscapes for ten paired random seeds, differing only in $\psi^{\text{BG}}$: flat ($\psi^{\text{BG}}=0$), the Gaussian landscape without conduits, and the same landscape with conduits.
The results are shown in Fig.~\ref{fig:control_summary}.

Both landscapes in which informed agents respond to $\psi^{\text{BG}}$ leave far fewer agents never exposed than the flat control (Fig.~\ref{fig:control_summary}a).
Relative to the flat baseline, the final susceptible count $S$ falls by $256.1\pm37.0$ agents without conduits and by $208.4\pm33.4$ agents with conduits (of 1000; Wilcoxon signed-rank, two-sided, $p=0.002$ for both comparisons).
As susceptible agents are effectively stationary (Sec.~\ref{sec:implementation}), this drop in $s$ is attributed to the sweeping of informed agents under $\psi^{\text{BG}}$, bringing susceptible agents into exposure that would not otherwise occur.

We note that guidance does not come without cost.
By $T=15$, conduits bring more agents to the safe zone than the unguided landscape (Fig.~\ref{fig:control_summary}b): $499.8\pm15.8$ agents (of 1000) lie within $0.2$ of $(0.5,0.5)$ with conduits, versus $408.9\pm15.7$ without (agents ever informed, i.e.\ currently informed or reverted: $75.0\pm2.6\%$ versus $57.4\pm2.1\%$; Appendix~\ref{sec:statistics}).
Yet conduits also leave $47.7\pm23.9$ more agents (of 1000) never exposed than the unguided landscape (Wilcoxon $p=0.004$, 9 of 10 seeds; Fig.~\ref{fig:control_summary}c).
An important consideration is the response window set by $\tau_I$, as reverted agents stop responding to $\psi^{\text{BG}}$ and essentially freeze in place.
Conduits shorten the evacuation trip and help an agent reach the safe zone before it reverts.
However, routing informed agents out of the domain faster and along a narrower path also reduces how much of the population those agents sweep past en route.
Consistent with this account, susceptible agents remaining at $T=15$ are disproportionately close to the hazards rather than far from them in both evacuating landscapes ($40.9\%$ and $34.7\%$ of remaining susceptibles lie within $0.15$ of a hazard centre without and with conduits, respectively, versus $18.1\%$ for flat), as expected if sweeping is concentrated along the route to the safe zone rather than at the hazards themselves.

We can account for the final exposed agent count as follows.
The multimodal mixture $M$ used here and in Sec.~\ref{sec:multimodal_study} draws 300 agents from a near-permanent-latency component ($\tau_E \sim \mathrm{Gamma}(\alpha,\theta), \alpha = 1000, \theta = 0.75$), which cannot leave the exposed state within $T=15$.
Because $\tau_E$ is assigned independently of an agent's position, this 300-agent share applies regardless of landscape, so the final exposed count should track $0.3$ times the number of agents ever exposed, $S_0 - s(T)$.
This prediction matches the observed $e(T)$ to within 7 agents in all three configurations (flat: $163.5\pm9.3$ predicted vs.\ $170.5\pm9.8$ observed; no conduits: $240.4\pm9.4$ vs.\ $246.6\pm14.1$; conduits: $226.1\pm10.7$ vs.\ $231.8\pm14.4$), confirming that the landscape governs $S(T)$ but leaves the latent fraction trapped within $e$ essentially unchanged, largely by construction (Fig.~\ref{fig:control_summary}c).

Separately, the contact process itself is below the two-dimensional continuum-percolation threshold: at the controls' contact radius $r=r_E+r_I=0.03$ and density $\rho=1000$, the expected number of neighbours within contact range is $\pi r^2\rho\approx2.8$, below the threshold of about $4.5$ for random discs \cite{mertensContinuumPercolationThresholds2012}. %
This bears on the static contact structure; the agents' own motion (Sec.~\ref{sec:st_dyn_wo_conduits}) is what allows exposure to exceed what a fixed network of this density would otherwise support.

\section{Discussion}
\label{sec:disc}

As demonstrated in our evacuation case studies, the coupling of environmental potentials with heterogeneous transition latencies generates highly localized, non-equilibrium spatial dynamics.
These emergent behaviors contrast sharply with the assumptions of classical mathematical models, which often treat populations as continuous and well-mixed.
When attempting to bridge these two modeling paradigms---specifically, by coarse-graining our discrete spatial model into traditional compartmental ODEs---we find that the physical realities of crowding and spatial correlation inherently violate standard mean-field assumptions.

\subsection{Effect of spatial heterogeneity on coarse-graining}

Classical compartmental models in epidemiology describe the ensemble-averaged evolution of populations transitioning between the $S,E,I,R$  states. Under a well-mixed approximation, these dynamics are typically represented by a system of nonlinear ordinary differential equations.
Transitions governed by state-independent waiting times, such as $E\to I$ and $I\to R$, naturally give rise to linear transition terms when the waiting times are exponentially distributed.
For example, exponentially distributed transition times with rates $\sigma$ and $\gamma$ yield rate terms of the form $\sigma E$ and $\gamma I$, respectively.

The proximity-based $S\to E$ transition is fundamentally different because its rate depends on interactions between agents in different compartments. In the classical SEIR model, this interaction is represented through:
\begin{equation}
\frac{dS}{dt}=-\beta S I,
\end{equation}
with a corresponding gain $\beta SI$ in the exposed population.
From the perspective of a spatial agent-based model, this expression can be interpreted as a mean-field closure of the underlying encounter process: $S$ represents the population available to transition, while $\beta I$ represents the effective rate at which a susceptible agent encounters an informed agent.
In the dilute, well-mixed limit, susceptible--informed encounters may be treated as approximately independent, providing a microscopic interpretation of the bilinear dependence on $S$ and $I$.

A constant $\beta$ assumes that the microscopic encounter dynamics can be represented by a single effective transition rate throughout the evolution of the system.
This assumption can be examined directly in our ABM results.
Over non-overlapping windows of $W$ time steps, we estimate
\begin{equation}
    \hat\beta_k = \frac{n_k/N}{\Delta t \sum_{t \in k} s(t)\, i(t)},
    \label{eq:beta_hat}
\end{equation}
where $n_k$ is the number of $S\to E$ transitions in window $k$.
As a reference, suppose agent positions were redrawn uniformly between proximity checks. A susceptible agent would then have a Poisson-distributed number of informed agents within $r = r_E + r_I$, with mean $\pi r^2 N i$ in the unit domain, and the estimator would return the well-mixed rate (see Appendix \ref{sec:well-mixed} for derivation):
\begin{equation}
    \beta_{\mathrm{wm}}(i) = \frac{1 - \exp\left(-\pi r^2 N i\right)}{\Delta t\, i}
    \;\longrightarrow\; \frac{\pi r^2 N}{\Delta t}
    \quad (i \to 0).
    \label{eq:beta_wm}
\end{equation}

\begin{figure}[tbp]
    \centering
    \includegraphics[width=\textwidth]{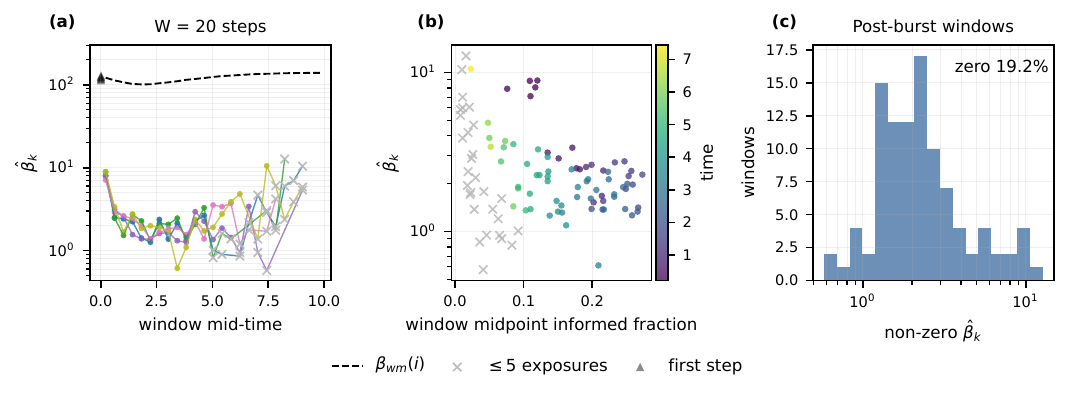}
    \caption{Effective $S\to E$ rate in flat-landscape simulations
    ($\psi^{\text{BG}} = 0$, $N = 1000$, $r_E = 0.03$, $r_I = 0$,
    $\tau_E \sim \mathrm{Gamma}(\alpha,\theta), \alpha = 1.5, \theta = 0.75$; $\Delta t = 0.02$, five seeds),
    estimated with Eq.~\eqref{eq:beta_hat} over windows of $W = 20$ steps.
    (a) $\hat\beta_k$ versus window mid-time for each seed. Triangles mark the first step, when agent positions are uniformly random; crosses mark windows with at most five exposures; the dashed curve is the well-mixed rate $\beta_{\mathrm{wm}}(i)$ of Eq.~\eqref{eq:beta_wm}.
    (b) $\hat\beta_k$ versus the informed fraction at the window midpoint, coloured by time.
    (c) Distribution of non-zero $\hat\beta_k$ after the first step; 19.2\% of these windows record no exposures.}
    \label{fig:beta_windows}
\end{figure}

Figure~\ref{fig:beta_windows} applies Eq.~\eqref{eq:beta_hat} to the flat-landscape simulations from Sec.~\ref{sec:temporal_heterogeneity}, with $r = 0.03$ and $\Delta t = 0.02$, for which $\pi r^2 N/\Delta t \approx 141$.
In the first step, when agent positions are uniformly random, $\hat\beta \approx 120$, in agreement with Eq.~\eqref{eq:beta_wm}.
Within the first time unit, however, $\hat\beta$ falls by a factor of about 60 and then fluctuates about $\hat\beta \approx 2$ for the remainder of the cascade (Fig.~\ref{fig:beta_windows}a).
Furthermore, windows with the same $i$ but at different times give different $\hat\beta$ (Fig.~\ref{fig:beta_windows}b), illustrating that the rate is not a function of the informed fraction alone.
A mean-field closure with constant $\beta$, or with $\beta(i)$, therefore misses the dominant effect, specifically the spatial correlations that the transitions themselves create.
Once an informed agent has exposed its susceptible neighbours, its neighbourhood is not replenished unless agents move appreciably relative to the contact radius, and further exposures proceed mainly through newly informed agents that are already adjacent to susceptible ones.

\begin{figure*}[t]
    \centering
    \includegraphics[width=\textwidth]{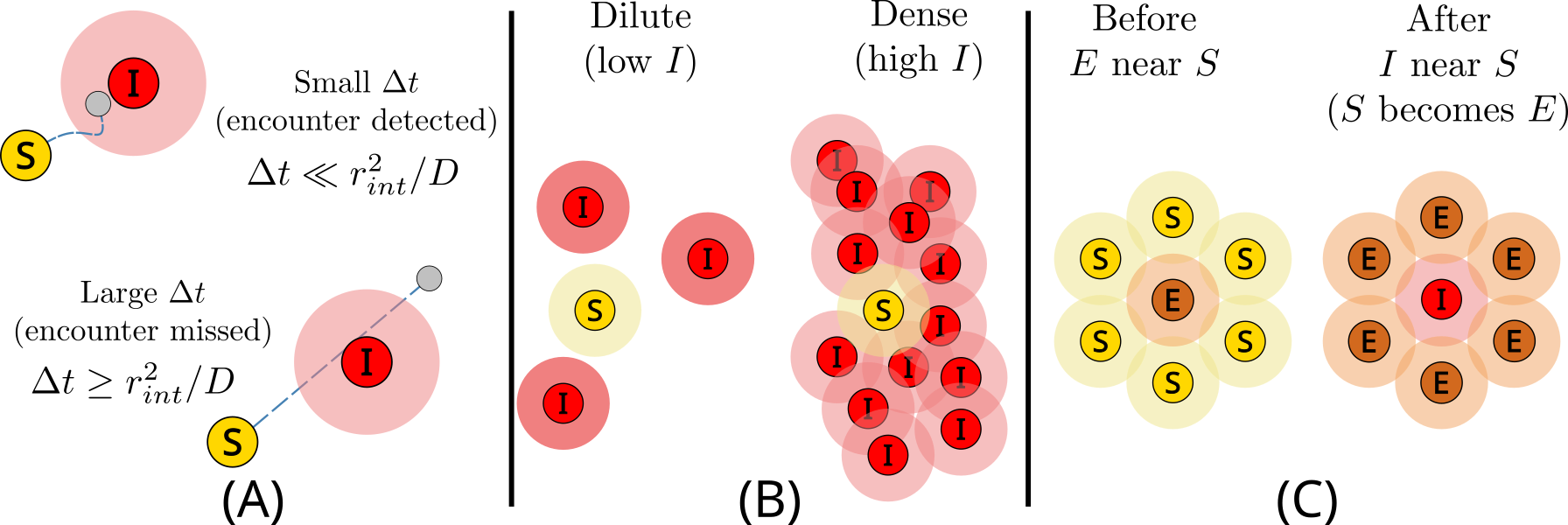}
\caption{
Microscopic mechanisms governing proximity-based transitions and their
coarse-grained representation.
\textbf{(A)} Effect of temporal resolution: for sufficiently small observation intervals, Brownian trajectories entering an informed agent's interaction neighborhood are resolved, whereas larger intervals can miss such encounters.
\textbf{(B)} Effect of spatial coverage: interaction neighborhoods are approximately independent in the dilute limit but increasingly overlap as the informed population grows, causing each additional informed agent to contribute
less new interaction area.
\textbf{(C)} Effect of spatial correlations inherited through state transitions: an exposed agent can transition to the informed state before substantial diffusive rearrangement, retaining susceptible neighbors within its interaction neighborhood and enabling subsequent local exposures.
}
    \label{fig:proximity_mechanisms}
\end{figure*}

A constant $\beta$ implicitly assumes that 1) individual encounters are adequately resolved in time, 2) interaction opportunities associated with different informed agents are approximately independent, and 3) spatial correlations generated by previous state transitions do not substantially influence subsequent encounters.
These assumptions define the dilute, well-mixed limit of the ABM.
Here, we show how the spatial dynamics depart from this limit through the three mechanisms below (Fig.~\ref{fig:proximity_mechanisms}).

\textbf{Temporal resolution of encounters.}
Brownian motion continuously changes the separation between agents, whereas proximity is evaluated at discrete observation times in the ABM.
The effective encounter rate therefore depends on how far agents typically diffuse between observations relative to the interaction distance.
When this displacement is small, proximity events are well resolved and the continuous-time encounter rate is recovered. At coarser temporal resolution, agents may traverse a significant fraction of the interaction neighborhood between observations, and the effective encounter rate can depend on the observation interval itself.
This effect is present even in the dilute limit and is therefore distinct from interactions between the neighborhoods of multiple informed agents.
Equation~\eqref{eq:beta_wm} makes this explicit: in the well-mixed limit each proximity check is an independent exposure opportunity, so the estimated rate scales as $1/\Delta t$ (see also Appendix~\ref{sec:timestep_effect}).

\textbf{Nonlinear spatial coverage.}
Even when individual encounters are well resolved, the contributions of different informed agents are approximately independent only in the dilute limit.
At low informed populations, interaction neighborhoods have little overlap, and each additional informed agent contributes approximately the same new region in which susceptible agents may be exposed.
As the informed population increases, however, these neighborhoods increasingly overlap.
Each additional informed agent then contributes progressively less new interaction area, causing the total region accessible for exposure to grow sub-linearly with the informed population.
The resulting transition rate therefore departs from the linear dependence on $I$ implicit in the classical bilinear ODE.

\textbf{Spatial correlations inherited through state transitions.}
Proximity-based transitions can themselves generate spatial correlations that persist as agents change compartment states.
This becomes particularly important when the $E\to I$ transition occurs on a timescale comparable to or shorter than that required for substantial diffusive rearrangement.
An exposed agent may then transition to the informed state while retaining approximately the same local neighborhood. Susceptible agents already located within its interaction range can subsequently be exposed without requiring a new encounter generated by Brownian motion.
The resulting local cascade retains information about the spatial configuration inherited from previous $S\to E\to I$ transitions.
Consequently, the transition dynamics may depend not only on the instantaneous compartment populations, but also on spatial correlations generated by their recent evolution.
The lack of a clear time dependence of $\hat\beta$ at fixed $i$ in Fig.~\ref{fig:beta_windows}b is a direct signature of this effect.

Together, these mechanisms highlight the assumptions underlying a constant-$\beta$ description of proximity-based transitions. Temporal resolution determines how individual Brownian encounters are observed; neighborhood overlap determines how the interaction opportunities generated by multiple informed agents accumulate; and state transitions can preserve spatial correlations that create additional exposure pathways. The classical $\beta SI$ law is recovered when these effects are negligible, whereas the spatial ABM generally leads to an emergent
transition rate that depends on both the encounter dynamics and the evolving spatial organization of the population.

\subsection{Implications for evacuation modeling}
\label{sec:implications}

Because this model has not been calibrated against any real-world evacuation data, its value here is not a direct recommendation for evacuation practice, but an approach to generate testable hypotheses about warning and evacuation design for further study.
The guidance trade-off of Sec.~\ref{sec:controls} follows from a sweeping mechanism: informed agents expose susceptible neighbors only while they remain near them, so guidance that gets informed agents to safety faster also shortens how long they spend sweeping past others who have not yet been exposed.
This suggests, as a hypothesis for further study, that keeping designated transmitters of information (e.g., wardens, fixed messengers) stationed near hazards rather than letting them evacuate as soon as they are informed could preserve the exposure benefit that guidance otherwise reduces.

A second hypothesis follows from the latency decomposition of Sec.~\ref{sec:controls}.
Since the final exposed count is set almost entirely by the fraction of the population with long response delays $\tau_E$ (independently of the landscape), no rearrangement of routes can recover agents who never leave the exposed state within the response window set by $\tau_I$.
This argues for interventions that shorten $\tau_E$ directly (e.g., direct, redundant alerts), rather than for investment in evacuation routes alone.

Both hypotheses connect to existing accounts of how people actually receive and act on warnings. Information often spreads informally, learned from neighbors and family rather than only official channels \cite{sorensenHazardWarningSystems2000}.
Furthermore, the pre-decisional stages of receiving, interpreting, and deciding to act, described by the Protective Action Decision Model \cite{lindellProtectiveActionDecision2012}, correspond to exactly the kind of delay $\tau_E$ represents at the agent level.

\subsection{Limitations and future work}
\label{sec:limitations}

As discussed in the previous section, we specifically highlight that the mechanistic modeling results presented here are not calibrated to evacuation data and make a number of simplifying assumptions.
We present the results in a 2D unit domain with a single hazard geometry (three Gaussian maxima + one central minimum); this may not generalize to other layouts without direct testing.
Our studies squarely fall in the drift-dominated regime (Sec.~\ref{sec:implementation}) where non-informed agents are effectively stationary, while $\varepsilon = 10^{-6}$ for the Lennard-Jones potential makes inter-agent forces (cohesion/repulsion) negligible in the case studies and controls (e.g., no crowd cohesion/jamming effects).
Our periodic boundaries, combined with a non-periodic potential, imply that agents at the boundaries experience a step force change.
Resolving pairwise interactions for large populations scales like an N-body problem; dynamic neighbor-list generation is a natural next optimization.

The flexibility of the linear superposition of potentials opens immediate avenues for future research opportunities.
This includes implementing dynamically evolving background potentials, such as the spatiotemporal propagation of a physical hazard, as well as introducing heterogeneous agent typologies driven by competing objective functions, allowing us to study the competitive dynamics of populations in the kinematics-information space.
The continuous, grid-free nature of the background potentials and transition functions is a modeling advantage to be exploited in future work.
Driving gradients $\nabla\psi$ are evaluated directly at each agent's coordinates rather than interpolated from a grid, avoiding the discretization artifacts of grid-based formulations while preserving mathematical tractability.
This naturally supports scaling to higher dimensions.
Additional dimensions can be utilized to represent continuous internal state variables---such as an agent's kinetic mobility, resource depletion, or dynamic susceptibility---allowing agents to traverse a complex internal state-dependent phase-space simultaneously alongside their physical navigation.
The time-continuous stochastic dynamics formulation naturally lends itself to analytic and numerical coarse-graining techniques, paving the way for future upscaling into macroscopic continuum models (e.g., Fokker-Planck formulations) where aggregate population density can be studied as a continuous field.

\section{Conclusion}
\label{sec:conclusion}

This paper  explores how  agent-based modeling with time delays can provide insight into how evacuations occur, in terms of both individual outcomes (the number of agents informed) and community outcomes (the number of agents reaching the safe zone).
Unlike instantaneous force-based models, which often result in an idealized, fluid-like evacuation, the introduction of Gamma-distributed processing delays ($\tau_E$) accounts for intrinsic transition lag.

Across the case studies (Sec.~\ref{sec:appl_numer}), informed motion itself spreads the warning.
Landscapes in which informed agents respond to hazards and safe zones leave far fewer agents never exposed than an otherwise identical flat control.
We note a striking outcome from our simulation examples: while the evacuation routes get more agents to the safe zone, they also leave more of them never exposed than unguided evacuation.
Thus, the trade-off shown exchanges population-level awareness for individual speed.
Independently, the effective transmission rate estimated from encounters between susceptible and informed agents falls roughly sixty-fold below its well-mixed value within the first time unit, because the transitions themselves generate spatial correlations that a mean-field closure with a constant or state-dependent transmission rate cannot capture.

Taken together, these results point to two factors for warning design that may be further explored with this framework (Sec.~\ref{sec:implications}): where information-carrying agents are stationed relative to hazards, and how quickly a population responds to a warning once received---the latter being the one that no improvement in routing can substitute for.

\section*{Ethics}

This work did not involve human participants, animals or personal data.

\section*{Data/Code accessibility statement}
All code, configuration files and the data needed to reproduce the figures will be made available at \url{https://github.com/mechanoChem/evacuation_dynamics}.%

\section*{Declaration of AI use}
AI-assisted large language models (Claude and Claude Code, Anthropic; Codex, OpenAI) were used throughout the development of this manuscript: drafting and revising text, aiding in data analyses, building simulation and plotting pipelines, and assisting with the development and debugging of the data-analysis scripts used to compute the reported statistics.
All simulation data, figures, and numerical results reported in this manuscript were produced by the authors' own simulation and analysis codebase; no data, figures, or results were generated by AI.
The authors reviewed all text and independently verified the reported numerical claims against the underlying simulation output.
The authors take full responsibility for the manuscript's content and conclusions.

\section*{Author Contributions}
B.J.: Methodology, Investigation, Data curation, Formal analysis, Software, Supervision, Writing - original draft, Writing - review \& editing;
L.D.: Investigation, Data curation, Formal analysis, Software, Writing - original draft, Writing - review \& editing;
R.G.: Investigation, Data curation, Formal analysis, Software, Writing - review \& editing;
S.S.: Conceptualization, Funding acquisition, Methodology, Project administration, Investigation, Software, Supervision, Writing - review \& editing;
K.G.: Conceptualization, Funding acquisition, Methodology, Project administration, Investigation, Supervision, Writing - review \& editing

\section*{Competing interests}
The authors declare no conflict of interest.

\section*{Funding}
This work was supported at the University of Southern California by DARPA under the COMPASS program via Agreement \#HR0011-25-3-0160. The scientific opinions expressed here reflect the authors' views, only.

\section*{Acknowledgements}
The authors appreciate and acknowledge helpful discussions with Dr. Reese Jones and Dr. Cosmin Safta, both at Sandia National Laboratories.
The authors acknowledge the Center for Advanced Research Computing (CARC) at the University of Southern California for providing computing resources that have contributed to the research results reported within this publication (URL: \url{https://carc.usc.edu}).

\bibliographystyle{unsrt}
\bibliography{references}

\appendix
\newpage

\section{Derivation of the agent drift}
\label{app:drift_derivation}

Here, we reduce the general configuration-dependent potential of Eq.~\ref{eq:agent_potential_general} to the form of Eq.~\ref{eq:psi_tot}, while deriving the resulting pairwise force.

For identical agents, the interaction contribution to $\psi_i$ can depend only on the positions of the other agents relative to agent $i$. With $\boldsymbol{r}_{ij}$ as in Eq.~\ref{eq:relative_position}, the potential written as
\begin{equation}
    \psi_i = \psi\big(\boldsymbol{X}_i;\,\{\boldsymbol{r}_{ij}\}_{j\neq i}\big),
\end{equation}
where the ordering of the relative positions is inconsequential.
The dependence on $\boldsymbol{X}_i$ alone describes the environment, and the dependence on $\{\boldsymbol{r}_{ij}\}$ describes the other agents.

We further assume that agent--agent interactions are pairwise and isotropic, so that the interaction between agents $i$ and $j$ depends only on their separation $r_{ij}$ and not on the orientation of $\boldsymbol{r}_{ij}$.
The potential then separates into a single background term and a sum of pair potentials,
\begin{equation}
    \psi_i = \psi_i^{\text{BG}}(\boldsymbol{X}_i) + \sum_{j\neq i}\phi(r_{ij}),
\end{equation}
which, with the state gating $\psi_i^{\text{BG}} = \chi_i\,\psi^{\text{BG}}$ and the cutoff $r_{ij}\leq r_c$, is Eq.~\ref{eq:psi_tot}.
Each pair term $\phi(r_{ij})$ is unchanged by translating or rotating the whole configuration, and the sum over $j$ is unchanged by relabelling the neighbors.

The force on agent $i$ follows from the gradient of $\psi_i$ with respect to $\boldsymbol{X}_i$. For the pair term, the chain rule and
\begin{equation}
    \nabla_{\boldsymbol{X}_i} r_{ij} = \frac{\boldsymbol{X}_i-\boldsymbol{X}_j}{r_{ij}}
\end{equation}
give the force exerted on agent $i$ by agent $j$,
\begin{equation}
    \boldsymbol{F}_{ij} = -\nabla_{\boldsymbol{X}_i}\phi(r_{ij}) = -\phi'(r_{ij})\,\frac{\boldsymbol{X}_i-\boldsymbol{X}_j}{r_{ij}},
\end{equation}
which is Eq.~\ref{eq:pairwise_force}.
Although $\phi$ depends only on the scalar separation, the resulting force acts along the line connecting the two agents, and $\boldsymbol{F}_{ji} = -\boldsymbol{F}_{ij}$.

Finally, the Wiener increments in Eq.~\ref{eq:cont-time-stoc} are independent across agents and satisfy
\begin{equation}
    \mathrm{d}\boldsymbol{W}_i \sim \mathcal{N}\big(\boldsymbol{0},\,\boldsymbol{I}\,\mathrm{d}t\big),
\end{equation}
so that the stochastic displacement $\sqrt{2D}\,\mathrm{d}\boldsymbol{W}_i$ has covariance $2D\,\boldsymbol{I}\,\mathrm{d}t$.

\section{Derivation of curvature factor and estimated effect}
\label{app:conduit_curvature}

\subsection{Conduit force on a curved centerline}
\label{app:conduit_curvature_derivation}

Near a conduit, we describe a point $\boldsymbol{X}$ by two coordinates: the arc length $s$ of its nearest centerline point $\boldsymbol{q}(s)$, and its signed distance $\ell$ from the centerline.
With $\boldsymbol{T}(s)$ the unit tangent and $\boldsymbol{N}(s)$ the unit normal pointing toward the inside of the bend,
\begin{equation}
    \boldsymbol{X} = \boldsymbol{q}(s) + \ell\,\boldsymbol{N}(s),
\end{equation}
so that $\boldsymbol{\delta} = \ell\,\boldsymbol{N}$ and $\|\boldsymbol{\delta}\| = |\ell|$.
Along a curve of curvature $\kappa$, the normal turns as $\mathrm{d}\boldsymbol{N}/\mathrm{d}s = -\kappa\,\boldsymbol{T}$.
Therefore,
\begin{equation}
    \mathrm{d}\boldsymbol{X} = (1-\kappa\ell)\,\boldsymbol{T}\,\mathrm{d}s + \boldsymbol{N}\,\mathrm{d}\ell.
\end{equation}
A step $\mathrm{d}s$ along the centerline  moves the point a distance $(1-\kappa\ell)\,\mathrm{d}s$: less on the inside of the bend ($\ell>0$) and more on the outside ($\ell<0$).
Projecting onto $\boldsymbol{T}$ and $\boldsymbol{N}$ gives $\nabla s = \boldsymbol{T}/(1-\kappa\ell)$ and $\nabla\ell = \boldsymbol{N}$, so by the chain rule, $\nabla\psi = (\partial\psi/\partial s)\,\nabla s + (\partial\psi/\partial\ell)\,\nabla\ell$, that is,
\begin{equation}
    \nabla\psi = \frac{\boldsymbol{T}}{1-\kappa\ell}\,\frac{\partial\psi}{\partial s} + \boldsymbol{N}\,\frac{\partial\psi}{\partial\ell},
\end{equation}
valid for $|\kappa\ell|<1$, that is, within one radius of curvature of the centerline.

For the conduit potential $\psi_{\text{conduit}} = -(A+g\,s)\exp(-B\ell^2)$,
\begin{equation}
    \frac{\partial\psi_{\text{conduit}}}{\partial s} = -g\,e^{-B\ell^2},
    \qquad
    \frac{\partial\psi_{\text{conduit}}}{\partial\ell} = 2B\ell\,(A+g\,s)\,e^{-B\ell^2},
\end{equation}
and the exact force is
\begin{equation}
    -\nabla\psi_{\text{conduit}}
    = \underbrace{-2B\,(A+g\,s)\,e^{-B\ell^2}\,\boldsymbol{\delta}}_{\boldsymbol{F}_{\text{restore}}}
    + \underbrace{\frac{g}{1-\kappa\ell}\,e^{-B\ell^2}\,\boldsymbol{T}}_{\boldsymbol{F}_{\text{forward}}}.
\end{equation}
The restoring term is Eq.~\ref{eq:channel_restoring_force} and the forward term is Eq.~\ref{eq:channel_forward_force}.
For a straight conduit ($\kappa=0$) or on the centerline ($\ell=0$), the forward term reduces to $g\,e^{-B\ell^2}\,\boldsymbol{T}$.

\subsection{Error from neglecting curvature}
\label{app:conduit_curvature_error}

The implementation omits the factor $1/(1-\kappa\ell)$ (Sec.~\ref{sec:implementation}), so the forward force it applies differs from the exact one by
\begin{equation}
    g\,e^{-B\ell^2}\left(\frac{1}{1-\kappa\ell}-1\right) = g\,e^{-B\ell^2}\,\frac{\kappa\ell}{1-\kappa\ell}.
\end{equation}
This error vanishes on the centerline, where $\ell = 0$, and far from it, where the Gaussian envelope vanishes.
It is largest at about one conduit width ($\ell\approx\sigma_c$) on the inside of the bend, and grows with curvature.
To evaluate this error, we need the curvature of each conduit and the conduit width.
The width follows from $B = 1/(2\sigma_c^2)$, so $\sigma_c = 1/\sqrt{2B}$; for $B = 400$, $\sigma_c \approx 0.035$.

Each conduit centerline is a quadratic Bezier curve from a starting point $\boldsymbol{P}_0$ to the safe zone $\boldsymbol{P}_2 = \boldsymbol{X}_{\text{min}}$,
\begin{equation}
    \boldsymbol{C}(u) = (1-u)^2\,\boldsymbol{P}_0 + 2u(1-u)\,\boldsymbol{P}_1 + u^2\,\boldsymbol{P}_2,
    \qquad 0 \le u \le 1.
\end{equation}
The control point $\boldsymbol{P}_1$ is offset from the midpoint $\boldsymbol{M} = (\boldsymbol{P}_0+\boldsymbol{P}_2)/2$ along the unit vector $\boldsymbol{e}_\perp$ perpendicular to $\boldsymbol{P}_2-\boldsymbol{P}_0$,
\begin{equation}
    \boldsymbol{P}_1 = \boldsymbol{M} \pm bL\,\boldsymbol{e}_\perp,
    \qquad L = \|\boldsymbol{P}_2-\boldsymbol{P}_0\|,
\end{equation}
where the bend parameter $b$ sets how far the conduit bows away from a straight line, and the sign sets the side.
The curvature of a parametric curve is $\kappa = \|\boldsymbol{C}'\times\boldsymbol{C}''\|/\|\boldsymbol{C}'\|^3$.
Here
\begin{equation}
    \boldsymbol{C}'(u) = (\boldsymbol{P}_2-\boldsymbol{P}_0) \pm 2bL(1-2u)\,\boldsymbol{e}_\perp,
    \qquad
    \boldsymbol{C}'' = \mp 4bL\,\boldsymbol{e}_\perp,
\end{equation}
so that $\|\boldsymbol{C}'\times\boldsymbol{C}''\| = 4bL^2$ and $\|\boldsymbol{C}'\|^2 = L^2\left[1 + 4b^2(1-2u)^2\right]$.
The curvature is therefore
\begin{equation}
    \kappa(u) = \frac{4b}{L\left[1 + 4b^2(1-2u)^2\right]^{3/2}},
\end{equation}
which is largest at the midpoint $u = 1/2$:
\begin{equation}
    \kappa_{\max} = \frac{4b}{L},
    \qquad
    R_{\min} = \frac{1}{\kappa_{\max}} = \frac{L}{4b}.
\end{equation}
All three conduits have $L = \|(0.4, 0.4)\| \approx 0.566$, so they differ only in $b$.

Finally, the error is largest near $\ell = \sigma_c$ on the inside of the sharpest bend, where $e^{-B\sigma_c^2} = e^{-1/2}$.
The largest error, as a fraction of the peak forward force $g$, is therefore approximately
\begin{equation}
    e^{-1/2}\,\frac{\kappa_{\max}\sigma_c}{1-\kappa_{\max}\sigma_c},
\end{equation}
which agrees with a numerical maximization over $\ell$ to within 0.3 percentage points for each conduit.
Table~\ref{tab:conduit_curvature} lists these quantities.
For the most strongly curved conduit, $\kappa_{\max}\sigma_c \approx 0.2$ and the error is about 15\% of $g$.
Beyond one radius of curvature, where the coordinates above fail, the Gaussian envelope $e^{-B\ell^2}$ is below $10^{-5}$.

\begin{table}[tbp]
\centering
\caption{Curvature of the three conduits and the largest error from omitting the factor $1/(1-\kappa\ell)$, as a fraction of the peak forward force $g$. Each conduit runs from the listed starting point to $\boldsymbol{X}_{\text{min}} = (0.5, 0.5)$, with $L \approx 0.566$ and $\sigma_c \approx 0.035$. $\kappa_{\max} = 4b/L$ and $R_{\min} = L/(4b)$ are attained at the midpoint of the centerline.}
\label{tab:conduit_curvature}
\begin{tabular}{lccccc}
\hline
Start & Bend $b$ & $\kappa_{\max}$ & $R_{\min}$ & $\kappa_{\max}\sigma_c$ & Largest error \\
\hline
$(0.1, 0.1)$ & 0.8 & 5.66 & 0.177 & 0.20  & 15\% \\
$(0.9, 0.9)$ & 0.6 & 4.24 & 0.236 & 0.15  & 11\% \\
$(0.1, 0.9)$ & 0.3 & 2.12 & 0.471 & 0.075 & 5\% \\
\hline
\end{tabular}
\end{table}

\section{Alternative conduit forcing}
\label{app:conduit_alt}
While this unified approach is mathematically conservative, it introduces a potentially undesirable physical artifact into the resulting kinematics: the magnitude of the lateral restoring force $\boldsymbol{F}_{\text{restore}}$ grows continuously with the arc-length $s(\boldsymbol{X})$.
Consequently, agents further down the channel would experience much stiffer lateral confinement than those near the entrance, artificially altering their local diffusion and interaction dynamics.
To eliminate this path-position dependence, one can apply the forward propulsion as an explicit, non-conservative forward driving field by setting $s(\boldsymbol{X}) = 0$.
By stripping the explicit path position $s(\boldsymbol{X})$ from the amplitude, the restorative cross-sectional potential becomes a function of lateral deviation alone.

\section{Parametric sensitivity analysis}
\label{sec:param_sens}

With the established framework, we proceed by establishing a baseline for how the system's parameters govern the population scale dynamics of information exchange.
By isolating key parameters---agent density, diffusivity, interaction radii, and boundary conditions---we observe their independent effects on the state-transition cascades: $S \to E \to I \to R$.
For these studies, the influence of the background potential is silenced by adopting a flat  landscape ($\psi^\text{BG}(\boldsymbol{X}) = 0$).
Periodic boundary conditions were imposed to maintain a conserved number of agents, $N$ (see Appendix~\ref{sec:bc_effect}).
Table \ref{tab:lammps_parameters} contains the baseline parameters used for these isolated studies.

\begin{table}[t]
\centering
\begin{tabular}{|l|l|}
\hline
\textbf{Description}                 & \textbf{Value/Expression}                 \\ \hline
Number of agents, $N$                     & $5 \times 10^{4}$                                    \\ \hline
Agents per unit area                 & $1 \times 10^4$                          \\ \hline
Informed agent fraction, $I$              & $1 \times 10^{-2}$                                     \\ \hline
Information radius, $r_I$                   & $4 \times 10^{-3}$                                    \\ \hline
Domain size, $l$                         & $\sqrt{\frac{n_{\text{agents}}}{\text{agents\_density}}}$ \\ \hline
Number of time steps, $n$                 & $3 \times 10^{3}$                                     \\ \hline
Time step, $\Delta t$                            & $5 \times 10^{-2}$                                     \\ \hline
Diffusivity $D$     & $1.38 \times 10^{-5}$                       \\ \hline
\end{tabular}
\caption{Default parameters used in the input files for this section.
}
\label{tab:lammps_parameters}
\end{table}

\subsection{Dependence on agent density}
\label{sec:density}

The frequency of spatial overlap between agents' spheres of interaction, and thus the rate of information transfer, is  constrained by spatial crowding.
We first evaluate the effect of varying the mean agent density $d = N/l^2$.

\begin{figure}[tbp]
    \centering
    \includegraphics[width=\linewidth]{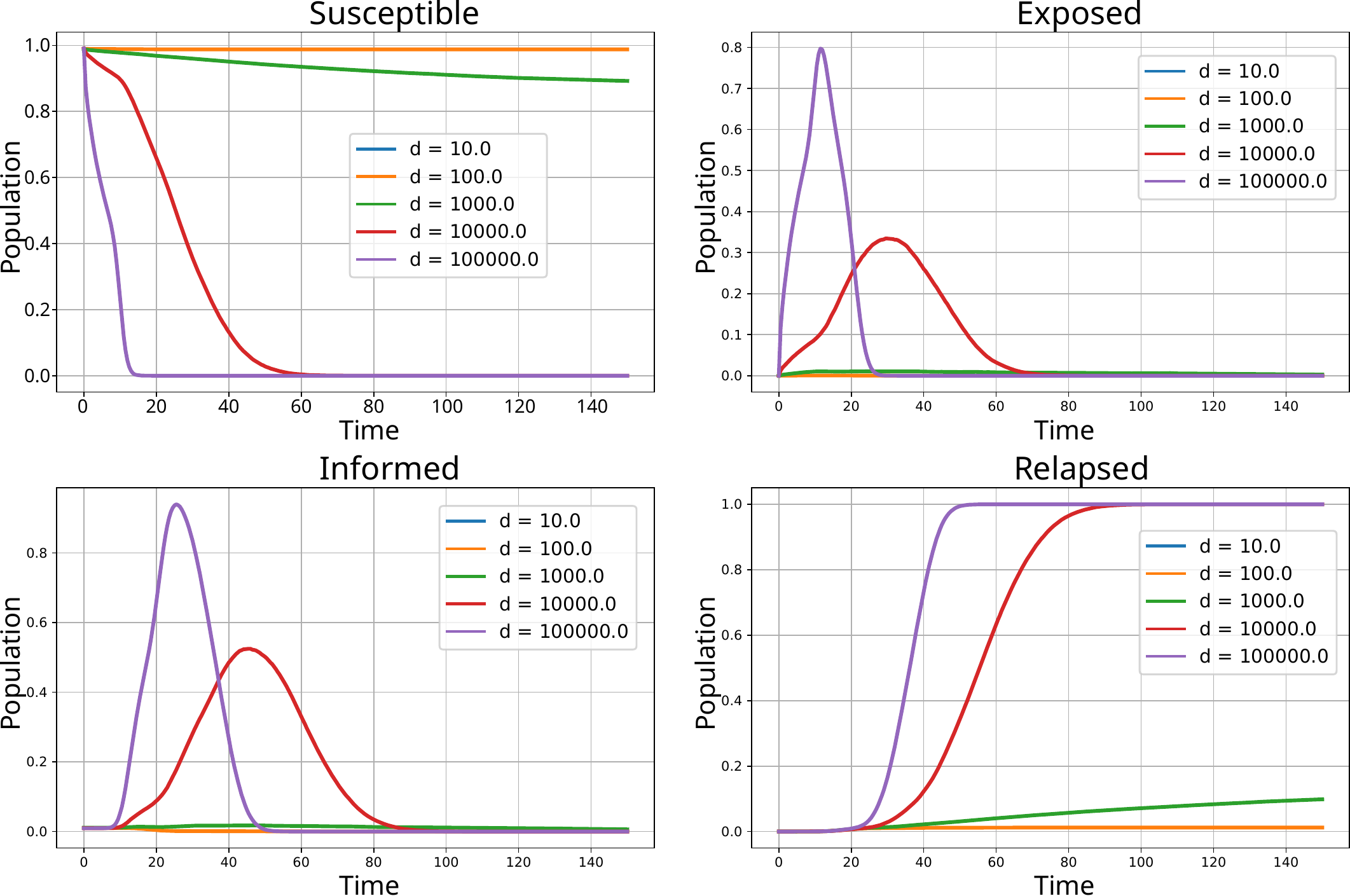}
    \caption{Evolution of total SEIR population fractions  for different densities. Starting fraction: $\{s,e,i,r\}=\{0.99,0,0.01,0\}$}
    \label{fig:case_study_density}
\end{figure}

Figure \ref{fig:case_study_density} demonstrates that as $d$ increases, the interactions between agents in different states grows, and there is a higher probability of overlapping spheres of interaction.
The dynamics are initiated by $S\to E$ transitions, which occur more frequently as $d$ increases.
Note that the overall flow is for agents that begin in the $S$ compartment to transition through $E$ and $I$, and terminate their transit in $R$.
Therefore, the susceptible fraction monotonically approaches $0$, while the exposed and informed fractions begin near $0$, reach intermediate maxima and eventually decline to $0$.
Also note that the informed population's rise and decline trails that of the exposed. Concurrently, the reverted fraction grows asymptotically toward $1.0$.

\subsection{Dependence on random agent movement (diffusivity)}
\label{sec:temp}

\begin{figure}[tbp]
    \centering
    \includegraphics[width=\linewidth]{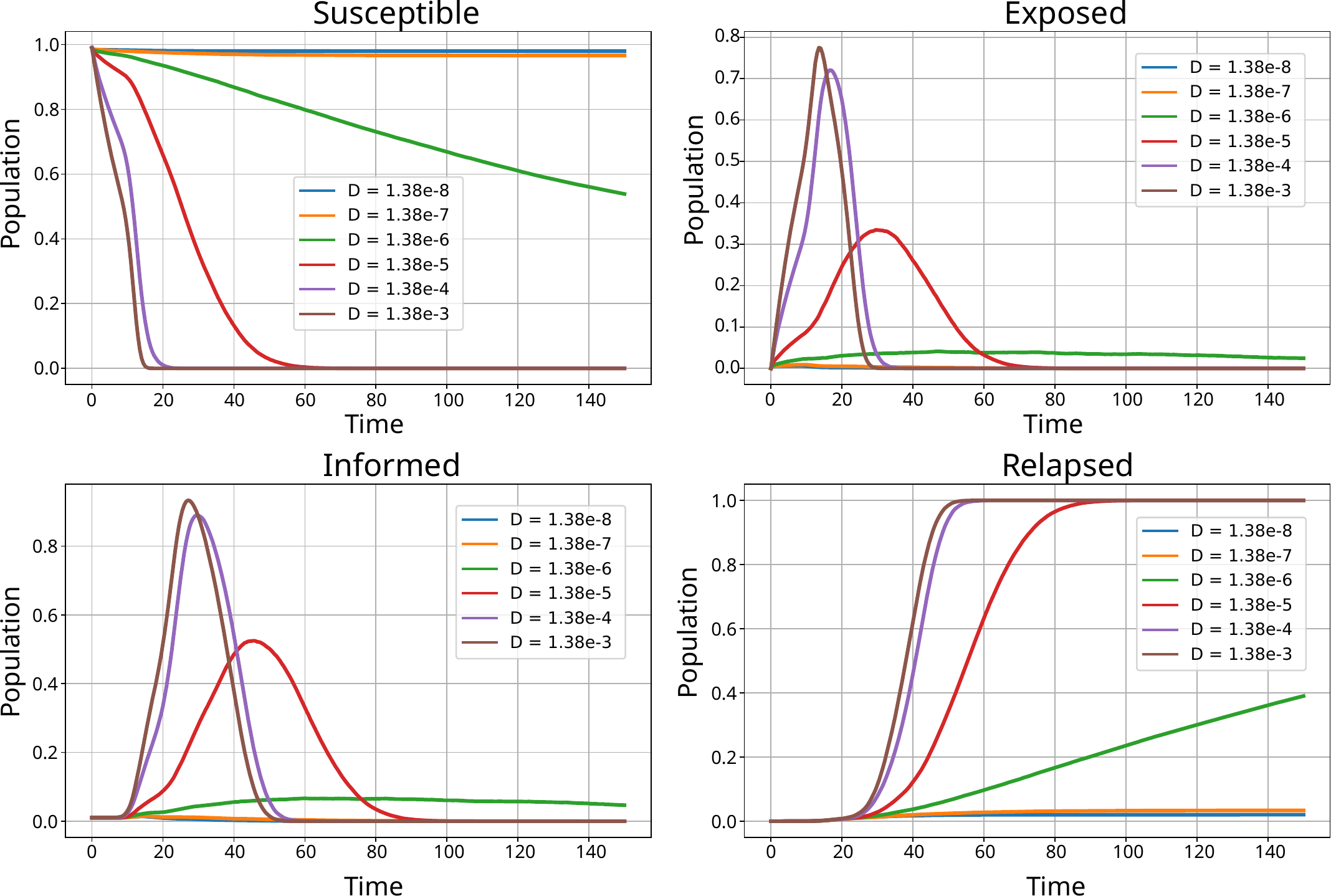}
    \caption{Evolution of total SEIR population fractions  for different diffusivities. Starting fraction: $\{s,e,i,r\}=\{0.99,0,0.01,0\}$}
    \label{fig:case_study_temperature}
\end{figure}

Increasing $D$  amplifies the  lengths of random walks and the frequency of collisions, which here is represented by overlaps of spheres of interaction.
Higher density and higher diffusivity therefore have similar emergent effects (Fig. \ref{fig:case_study_temperature}).

\subsection{Dependence on information radius}

\begin{figure}[tbp]
    \centering
    \includegraphics[width=\linewidth]{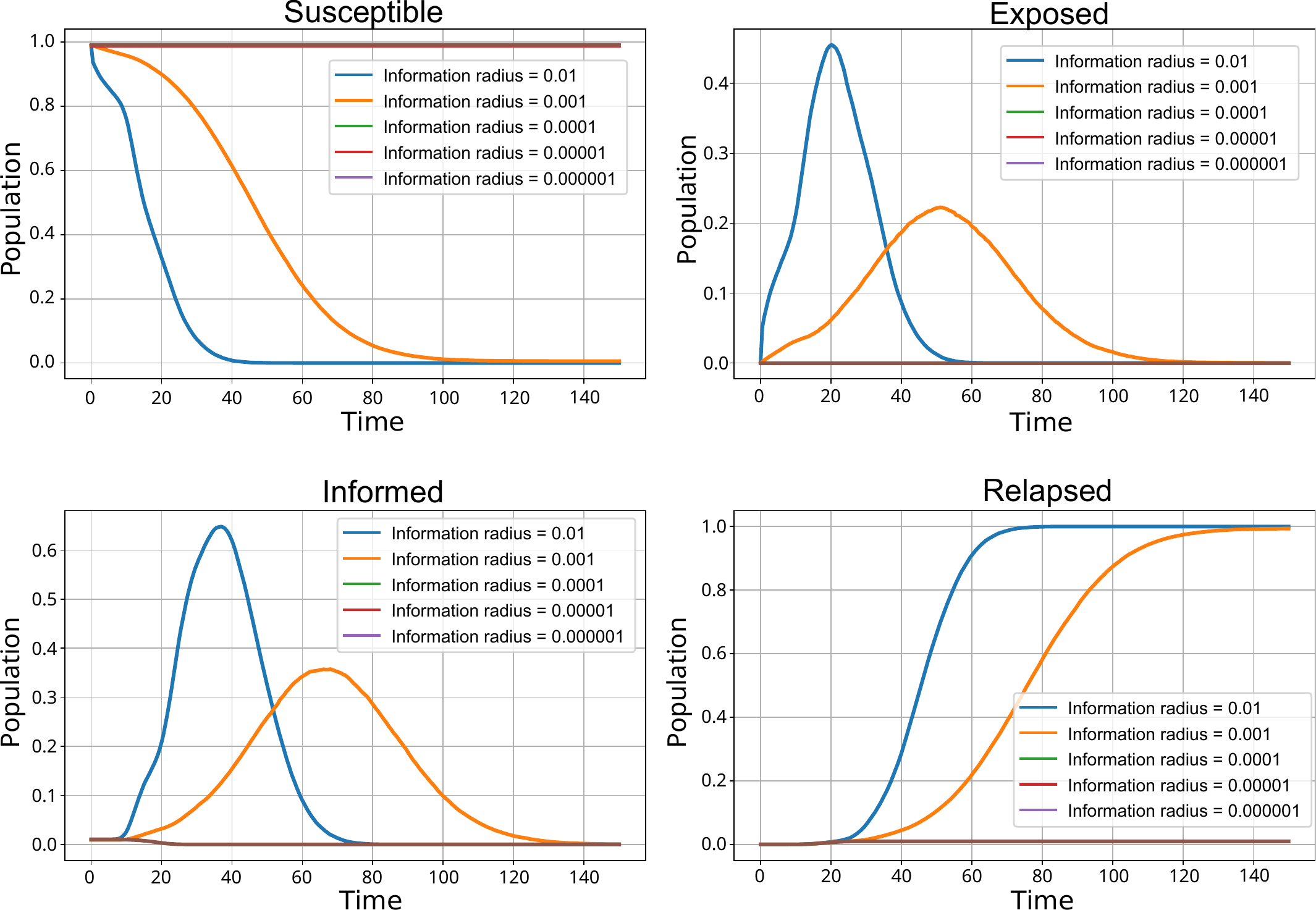}
    \caption{Evolution of total SEIR population fractions  for different information radii. Starting fraction: $\{s,e,i,r\}=\{0.99,0,0.01,0\}$}
    \label{fig:case_study_infection_radius}
\end{figure}

The information transmission cascade is naturally sensitive to the information radius $r_I$, which directly translates proximity to transmission.
Because $r_I$ is independent of the Lennard-Jones length scale $\sigma$, the communication range can be varied without changing the inter-agent spacing.
Larger information radii $r_I$ allow interactions between $S,I$ agent pairs over a larger neighborhood, thus boosting the cascade of dynamics: $S\to E\to I\to R$ (Fig. \ref{fig:case_study_infection_radius}).

\subsection{Dependence on boundary condition}
\label{sec:bc_effect}
\begin{figure}[tbp]
    \centering
    \includegraphics[width=\linewidth]{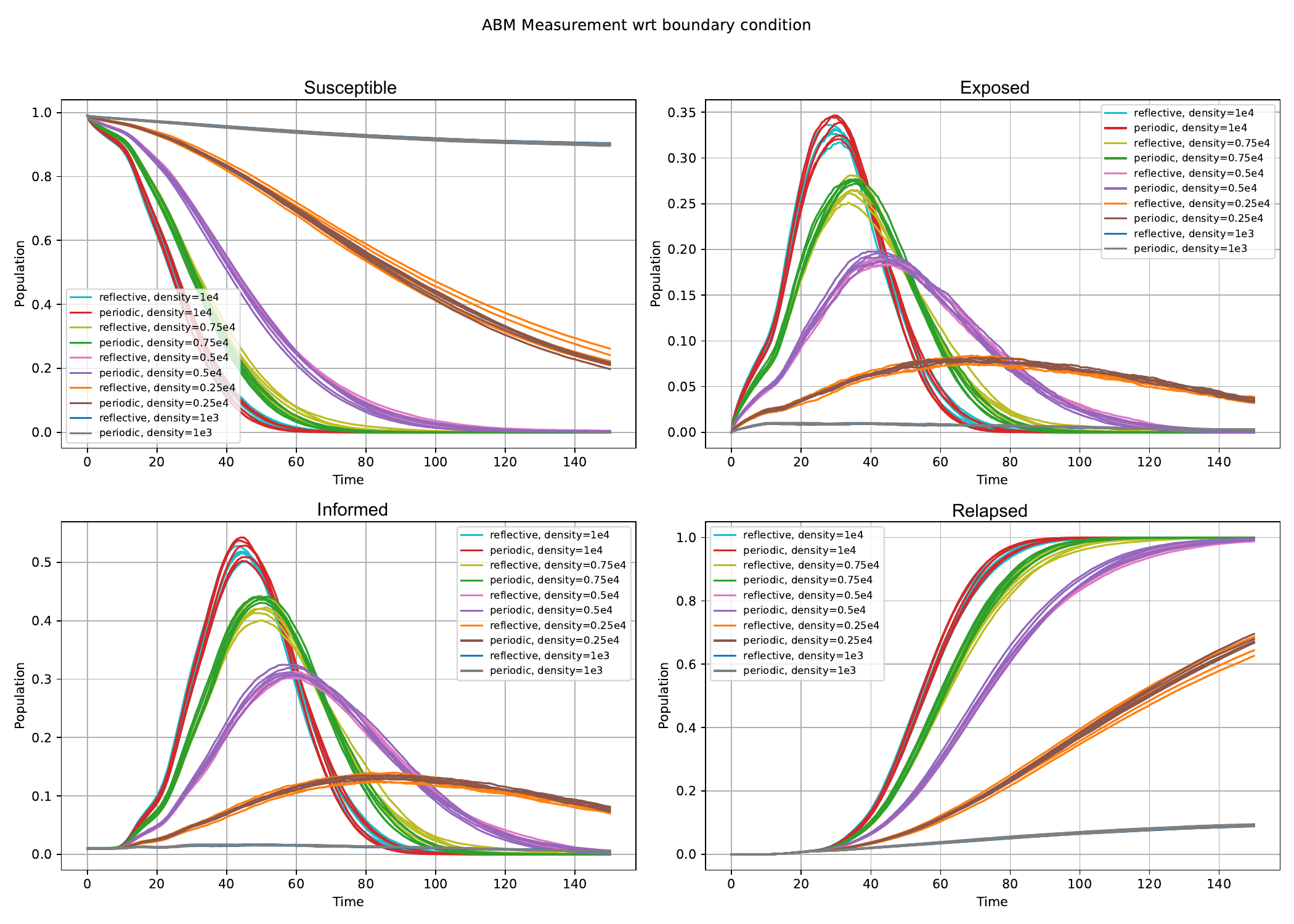}
    \caption{SEIR compartment fraction evolution for different boundary conditions. Starting fraction: $\{s,e,i,r\}=\{0.99,0,0.01,0\}$}
    \label{fig:case_study_boundary_condition}
\end{figure}

Figure \ref{fig:case_study_boundary_condition} compares both periodic and reflective boundary conditions in the simulation.
The difference between periodic and reflective boundary conditions is minimal for densities in the range $[10^3,10^4]$.
Any variations in dynamics are due to initial conditions of agent positions, which are randomized.

\section{Simulation parameters and computational details}
\label{app:sim_params}

Neighbor relationships were computed at each timestep\footnote{The effect of timestep is explored in Appendix \ref{sec:timestep_effect}} by first assembling agent positions into a NumPy array and evaluating the full pairwise Euclidean distance matrix using array broadcasting.
Distances exceeding a prescribed cutoff radius were set to zero, and the thresholded matrix was converted to SciPy’s \cite{virtanenSciPy10Fundamental2020} compressed sparse row (CSR) format.
The resulting CSR matrix provides both compact storage of retained local interactions and efficient access to nonzero neighbor pairs.
These sparse pairs are reused for the state update, where susceptible exposure is evaluated by comparing retained pair distances against the sum of the susceptible agent’s exposure radius $r_E$ and the informed neighbor’s information radius $r_I$.
Thus, subsequent interaction and transmission calculations scale with the number of retained neighbor pairs; however, the implementation still forms the dense all-pairs distance matrix before cutoff, so the neighbor-search stage itself remains quadratic in the number of agents.

Table~\ref{tab:case_params} summarizes the parameters of every simulation reported in the main text.

\begin{table}[tbp]
\centering
\caption{Simulation parameters. $\Gamma(\alpha,\theta)$ denotes a Gamma distribution with shape $\alpha$ and scale $\theta$. Delay distributions: S, $\Gamma(1.5, 0.75)$; B$_2$, 70\% $\Gamma(1.5, 0.75)$ and 30\% $\Gamma(4.5, 0.75)$; B$_3$, 70\% $\Gamma(1.5, 0.75)$ and 30\% $\Gamma(1000, 0.75)$; M, 50\% $\Gamma(1.5, 0.75)$, 20\% $\Gamma(4.5, 0.75)$ and 30\% $\Gamma(1000, 0.75)$. Landscapes: G, Gaussian hazards and safe zone; G+C, with conduits. In all runs the domain is $\Omega = (0,1)^2$ with periodic boundaries and $N = 1000$ agents start at uniform random positions; initial states are 900 susceptible and 100 informed (in the delay-study runs, drawn with probabilities 0.9 and 0.1). The update times are evenly spaced including both endpoints, so $\Delta t = T/(n_t - 1)$ (values rounded). The noise enters the update as $\sigma_\eta\sqrt{\Delta t}\,\boldsymbol{\eta}$, so $D = \sigma_\eta^2/2 = 5\times10^{-7}$ in every run. The delay-study runs used an earlier code version, with a stronger Lennard-Jones interaction.
$^{a}$Exposure is evaluated only for pairs within $r_c$, so the effective contact radius is $0.1$.}
\label{tab:case_params}
\small
\setlength{\tabcolsep}{4pt}
\begin{tabular}{llllll}
\hline
 & Fig.~\ref{fig:bimodal_tau_i} & Fig.~\ref{fig:multimodal_tau_e} & Cases 1--2 & Case 3 & Controls \\
\hline
$T$ ($n_t$)                & 4 (150)                     & 10 (500)                    & 15 (1500)                    & 15 (1500)            & 15 (1500) \\
$\Delta t$                 & $0.027$                     & $0.020$                     & $0.010$                      & $0.010$              & $0.010$ \\
$\sigma_\eta$              & $10^{-3}$                   & $10^{-3}$                   & $10^{-3}$                    & $10^{-3}$            & $10^{-3}$ \\
LJ $\varepsilon$           & $10^{-4}$                   & $10^{-4}$                   & $10^{-6}$                    & $10^{-6}$            & $10^{-6}$ \\
LJ $\sigma$                & $2^{-1/6}\times10^{-2}$     & $2^{-1/6}\times10^{-2}$     & $10^{-3}$                    & $10^{-3}$            & $10^{-3}$ \\
Cutoff $r_c$               & $0.1$                       & $0.1$                       & $0.1$                        & $0.1$                & $0.1$ \\
$r_E$ / $r_I$              & $0.03$ / $0$                & $0.03$ / $0$                & $0.3^{a}$ / $0$              & $0.03$ / $0$         & $0.03$ / $0$ \\
$\tau_E$                   & S or B$_2$                  & S, B$_3$ or M               & $\Gamma(10^{-5}, 0.75)$      & M                    & M \\
$\tau_I$                   & S or B$_2$                  & S                           & $\Gamma(800, 0.75)$          & $\Gamma(14, 0.75)$   & $\Gamma(14, 0.75)$ \\
$\psi^{\text{BG}}$         & none                        & none                        & G; G+C                       & G; G+C               & none; G; G+C \\
\hline
\end{tabular}
\end{table}

\section{Probability density functions}
\label{sec:pdf}

\begin{figure}
  \centering
  \begin{subfigure}[b]{0.5\textwidth}
    \centering
    \includegraphics[width=\linewidth]{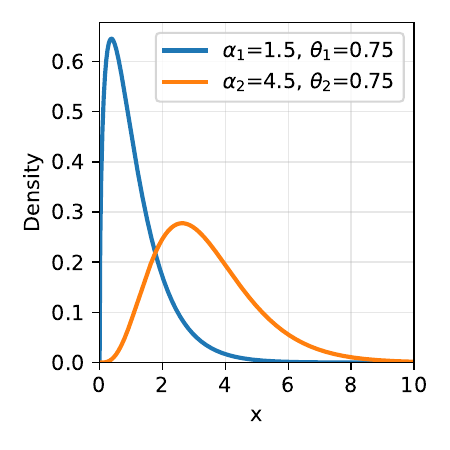}
  \end{subfigure}\hfill
  \begin{subfigure}[b]{0.5\textwidth}
    \centering
    \includegraphics[width=\linewidth]{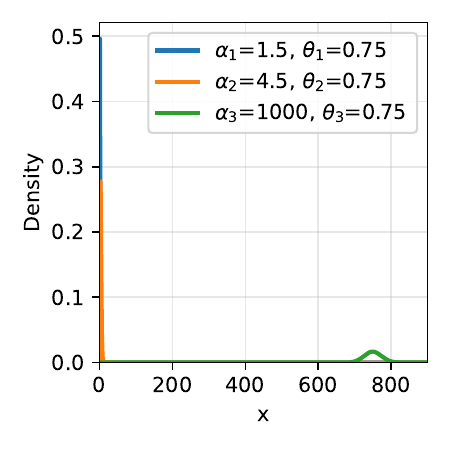}
  \end{subfigure}
  \caption{Left: Probability density function (PDF) of $\tau$ for two different shape, $\alpha$, and scale, $\theta$, values; typically labeled as the ``bimodal'' settings in this work. Right: Additional curve highlighting the effect of a large $\alpha$ parameter.}
  \label{fig:pdf}
\end{figure}

\section{Control experiment - statistical analysis}
\label{sec:statistics}

Table~\ref{tab:control_final} lists the final ($T=15$) agent counts and safe-zone fractions for the three control landscapes of Sec.~\ref{sec:controls}, over $n=10$ paired seeds $k=1,\dots,n$.
For each seed, $s_k$, $e_k$, $i_k$ and $r_k$ are the numbers of agents (of $N=1000$) in each state at $T$.
The safe zone is $Z=\{\boldsymbol{X} : d(\boldsymbol{X},\boldsymbol{X}_c)\le 0.2\}$ with $\boldsymbol{X}_c=(0.5,0.5)$, where $d$ is the Euclidean distance on $\Omega$:
\begin{equation}
    d(\boldsymbol{X},\boldsymbol{Y}) = \big\|\boldsymbol{X}-\boldsymbol{Y}\big\|.
\end{equation}
With $\boldsymbol{X}_j(T)$ the final position of agent $j$ and $\mathcal{I}_k$ and $\mathcal{R}_k$ the sets of informed and reverted agents at $T$, the fractions of all agents and of ever-informed agents in the safe zone are
\begin{equation}
    f_k^{\text{all}} = \frac{1}{N}\sum_{j=1}^{N}\mathbf{1}\big[\boldsymbol{X}_j(T)\in Z\big], \qquad
    f_k^{I+R} = \frac{1}{i_k+r_k}\sum_{j\in\mathcal{I}_k\cup\mathcal{R}_k}\mathbf{1}\big[\boldsymbol{X}_j(T)\in Z\big].
\end{equation}
Reverted agents are included in $f^{I+R}$.
Every reverted agent was once informed, and most agents informed early, including the 100 initially informed, have reverted by $T=15$ ($\tau_I\sim\Gamma(14,0.75)$, mean $10.5$).
Reverted agents no longer respond to $\psi^{\text{BG}}$ and stay where they reverted, so a reverted agent counts as in the safe zone only if it reached the zone while informed.
For each quantity $x_k$, the table reports the seed mean and sample standard deviation,
\begin{equation}
    \bar{x} = \frac{1}{n}\sum_{k=1}^{n}x_k, \qquad
    \mathrm{sd}(x) = \sqrt{\frac{1}{n-1}\sum_{k=1}^{n}\left(x_k-\bar{x}\right)^2}.
    \label{eq:seed_mean_sd}
\end{equation}

\begin{table}[!htbp]
\centering
\caption{Final state of the control experiment at $T=15$, seed mean $\pm$ sample s.d.\ over $n=10$ paired seeds. Counts are out of $N=1000$ agents. Safe-zone fractions $f^{\text{all}}$ and $f^{I+R}$ are given in percent.}
\label{tab:control_final}
\small
\setlength{\tabcolsep}{3pt}
\begin{tabular}{@{}lllllll@{}}
\hline
 & \multicolumn{4}{c}{Final count} & \multicolumn{2}{c}{Safe zone (\%)} \\
Landscape   & $s$              & $e$              & $i$              & $r$              & $f^{\text{all}}$   & $f^{I+R}$ \\
\hline
Flat        & $354.9\pm30.8$   & $170.5\pm9.8$    & $152.3\pm17.5$   & $322.3\pm31.2$   & $12.96\pm0.61$     & $14.5\pm2.3$ \\
No conduits & $98.8\pm31.2$    & $246.6\pm14.1$   & $171.7\pm14.4$   & $482.9\pm27.7$   & $40.89\pm1.57$     & $57.4\pm2.1$ \\
Conduits    & $146.5\pm35.6$   & $231.8\pm14.4$   & $158.6\pm14.5$   & $463.1\pm32.2$   & $49.98\pm1.58$     & $75.0\pm2.6$ \\
\hline
\end{tabular}
\end{table}

The landscapes and their seed-paired differences are compared using a two-sided Wilcoxon signed-rank test (Table~\ref{tab:control_wilcoxon}).
The same seed $k$ is used in every landscape.
For two landscapes $A$ and $B$ and a final-state quantity $x$ (either $s$ or $s+e$), the paired difference is
\begin{equation}
    d_k = x_k^{A} - x_k^{B}, \qquad k=1,\dots,n,
\end{equation}
with a mean and s.d. (Eq.~\eqref{eq:seed_mean_sd}).
A negative $d_k$ means that, for seed $k$, landscape $A$ leaves fewer agents never exposed ($x=s$) or never informed ($x=s+e$) than $B$.
Although seeds with $d_k=0$ would be dropped, this did not occur.
Let $R_k$ be the rank of $|d_k|$ among $|d_1|,\dots,|d_n|$. %
Any tied values receive their average rank.
The positive and negative rank sums are
\begin{equation}
    W^{+} = \sum_{k:\,d_k>0} R_k, \qquad W^{-} = \sum_{k:\,d_k<0} R_k, \qquad W^{+}+W^{-} = \frac{n(n+1)}{2} = 55,
\end{equation}
and the test statistic is $W=\min(W^{+},W^{-})$. %
The null hypothesis is that distribution of $d_k$ is symmetric about zero, so each $d_k$ is equally likely to be positive or negative.
As a result, all $2^{n}$ sign assignments to the ranks are equally likely. %
The exact two-sided $p$-value is
\begin{equation}
    p = \frac{1}{2^{n}}\,\#\big\{\text{sign assignments with } \min(W^{+},W^{-}) \le W_{\text{obs}}\big\}.
\end{equation}
The $p$-value is the probability that random signs alone would give a split of rank sums at least as lopsided as the one observed.
With $n=10$, $W=0$ requires every difference to have the same sign.
Only two of the $2^{10}=1024$ assignments (all positive or all negative) achieve this, so $p=2/1024\approx0.0020$, the smallest $p$ attainable with ten seeds. This holds even when ranks are tied.
$W=1$ additionally allows the smallest $|d_k|$ to take the opposite sign, giving $p=4/1024\approx0.0039$.
The values were computed with \texttt{scipy.stats.wilcoxon} \cite{virtanenSciPy10Fundamental2020} using its exact distribution.

\begin{table}[!htbp]
\centering
\caption{Paired Wilcoxon signed-rank tests between control landscapes ($n=10$ paired seeds, two-sided, exact). The difference is $d_k = x_k^{A}-x_k^{B}$ for comparison $A-B$, reported as mean $\pm$ sample s.d.\ in agents (of $N=1000$); $n_{+}$ is the number of seeds with $d_k>0$.}
\label{tab:control_wilcoxon}
\small
\setlength{\tabcolsep}{4pt}
\begin{tabular}{@{}llllll@{}}
\hline
Comparison $A-B$          & $x$     & $\bar{d}\pm\mathrm{sd}(d)$ & $n_{+}/n$ & $W$ & $p$ \\
\hline
No conduits $-$ Flat      & $s$     & $-256.1\pm37.0$           & $0/10$    & $0$ & $0.0020$ \\
                          & $s+e$   & $-180.0\pm26.2$           & $0/10$    & $0$ & $0.0020$ \\
Conduits $-$ Flat         & $s$     & $-208.4\pm33.4$           & $0/10$    & $0$ & $0.0020$ \\
                          & $s+e$   & $-147.1\pm22.3$           & $0/10$    & $0$ & $0.0020$ \\
Conduits $-$ No conduits  & $s$     & $47.7\pm23.9$             & $9/10$    & $1$ & $0.0039$ \\
                          & $s+e$   & $32.9\pm16.4$             & $9/10$    & $1$ & $0.0039$ \\
\hline
\end{tabular}
\end{table}

\section{Well-mixed rate calculation}
\label{sec:well-mixed}

We evaluate $\beta$ under a well-mixed assumption, starting from its definition:
\begin{equation}
    \beta = \frac{- \dot{s}}{si}
\end{equation}
where $\dot{s}$ is the rate of change of the susceptible fraction, while $s$ and $i$ are the fractions of susceptible and informed agents, $s, i \in [0,1]$.
To begin, redraw all positions uniformly in the unit periodic domain before every proximity check.
A susceptible agent will be exposed if at least one informed agent lies within $r=r_E + r_I$, with a contact disk of area $p = \pi r^2$ (since the domain has unit area, this is also the probability that a uniformly placed agent lands inside it).
We have $M = Ni$ informed agents.
The chance that a single informed agent misses the disk is $1-p$; assuming independence, the chance that all informed agents miss it is $(1-p)^M$.
Conversely, the chance that at least one hits the disk and the susceptible agent is exposed is $P=1-(1-p)^M$.

For a small contact disk ($p \ll 1$), we can approximate $(1-p) \approx e^{-p}$.
This allows us to do the following:

\begin{equation}
    (1-p)^M \approx e^{-pM} = e^{-\lambda},
    \qquad
    \lambda = pM = \pi r^2 N i
\end{equation}
\begin{equation}
    P \approx 1-e^{-\lambda}
\end{equation}
There are $sN$ susceptible agents, each exposed with probability $P$, so the expected number of exposures per step is $sNP$.
As a fraction of the population, $-\Delta s = sP$ per step, so $-\dot{s} \approx sP/\Delta t$.
This leads to:
\begin{equation}
    \beta_\text{wm} =
    \frac{-\dot s}{si} =
    \frac{sP/\Delta t}{si} = \frac{P}{\Delta t\, i} =
    \frac{1-e^{-\pi r^2 N i}}{\Delta t\, i}
\end{equation}

As $i \to 0$ ($\lambda \ll 1$), we can approximate $1-e^{-\lambda} \approx \lambda$.
Therefore,
\begin{equation}
    \beta_{\text{wm}}
    \approx
    \frac{\lambda}{\Delta t\, i} =
    \frac{\pi r^2 N i}{\Delta t\, i} =
    \frac{\pi r^2 N}{\Delta t}
\end{equation}

\section{Dependence on timestep}
\label{sec:timestep_effect}

\begin{figure}[tbp]
    \centering
    \includegraphics[width=\linewidth]{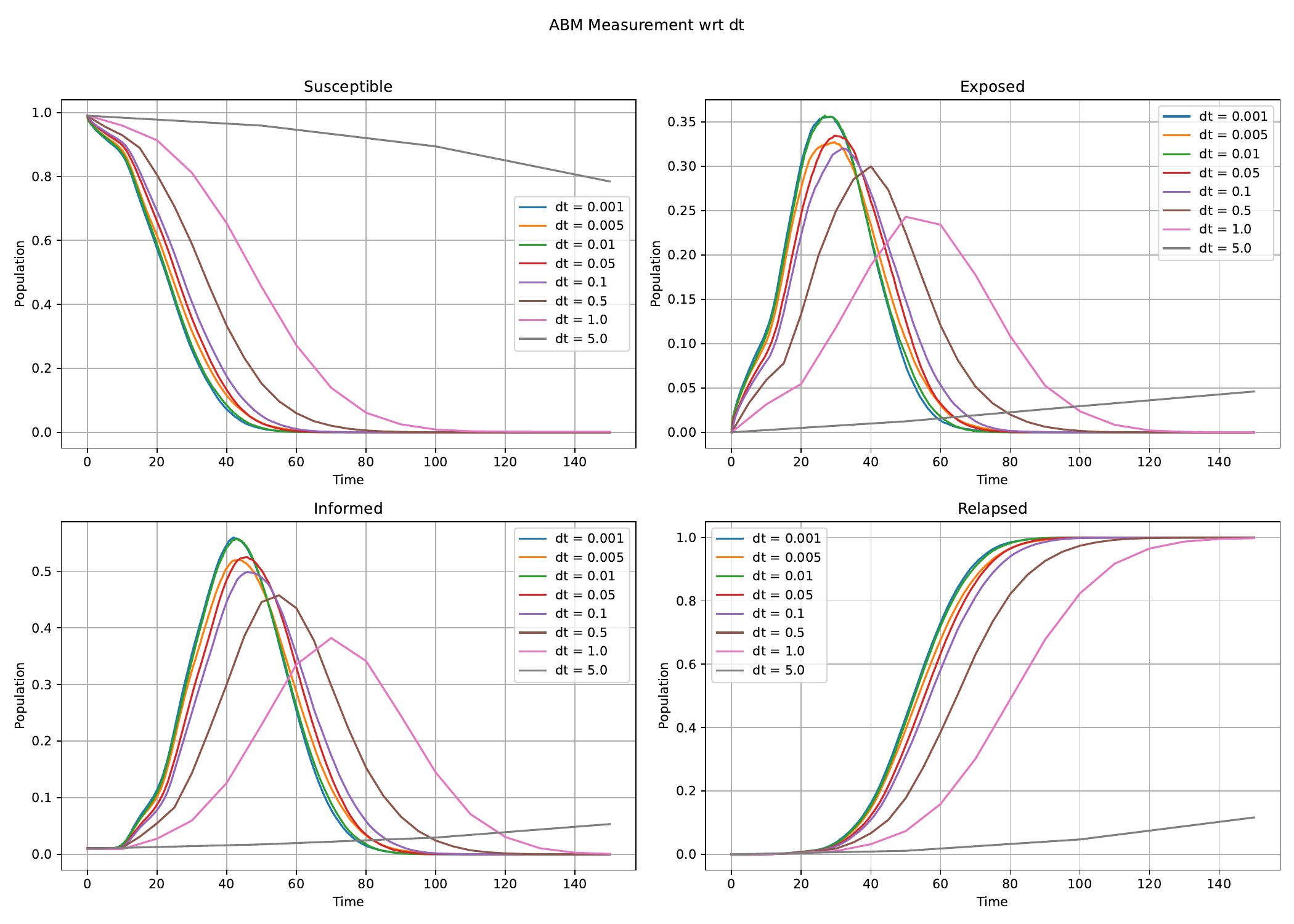}
    \caption{SEIR compartment fraction evolution for different timesteps. Starting fraction: $\{s,e,i,r\}=\{0.99,0,0.01,0\}$}
    \label{fig:case_study_dt}
\end{figure}

Figure \ref{fig:case_study_dt} explores the effect of timestep on state evolution.
At finer time steps, $\Delta t$, more configurations are explored over a given time interval. This leads to more transitions $S\to E$, allowing the remainder of the cascade to unfold: $E\to I\to R$.
Because exposure is checked at discrete times, the number of $S\to E$ transitions depends on $\Delta t$ (Remark~1). The results in the main text are therefore reported at the stated $\Delta t$ of each study rather than in a $\Delta t \to 0$ limit.

\end{document}